\documentclass[english,11pt, draftcls, technote, a4paper, oneside, onecolumn]{IEEEtran}
\usepackage[T1]{fontenc}
\usepackage[latin1]{inputenc}
\usepackage{bm}
\usepackage{amsmath}
\usepackage{amssymb}
\usepackage{graphicx}

\makeatletter
  \newtheorem{problem}{Problem}
  \newtheorem{defQED}{Definition}
  \newenvironment{lyxDefQED}{\begin{defQED}}{~\hfill\IEEEQEDclosed\end{defQED}}
  \newtheorem{thmQED}{Theorem}
  \newenvironment{lyxThmQED}{\begin{thmQED}}{~\hfill\IEEEQEDclosed\end{thmQED}}
  \newtheorem{lemQED}{Lemma}
  \newenvironment{lyxLemQED}{\begin{lemQED}}{~\hfill\IEEEQEDclosed\end{lemQED}}
  \newtheorem{remrk}{Remark}
  \newtheorem{corQED}{Corollary}
  \newenvironment{lyxCorQED}{\begin{corQED}}{~\hfill\IEEEQEDclosed\end{corQED}}

\usepackage{color}

\makeatother

\usepackage{babel}
\begin{document}

\title{Convergence Analysis of Saddle Point Problems in Time Varying Wireless
Systems - Control Theoretical Approach}

\author{Junting Chen and Vincent K. N. Lau}
\maketitle
\begin{abstract}
Saddle point problems arise from many wireless applications, and primal-dual
iterative algorithms are widely applied to find the saddle points.
In the existing literature, the convergence results of such algorithms
are established assuming the problem specific parameters remain unchanged
during the iterations. However, this assumption is unrealistic in
time varying wireless systems, as explicit message passing is usually
involved in the iterations and the channel state information (CSI)
may change in a time scale comparable to the algorithm update period.
This paper investigates the convergence behavior and the tracking
error of primal-dual iterative algorithms under time varying CSI.
The convergence results are established by studying the stability
of an equivalent virtual dynamic system derived in the paper, and
the Lyapunov theory is applied for the stability analysis. We show
that the average tracking error is proportional to the time variation
rate of the CSI. Based on these analyses, we also derive an adaptive
primal-dual algorithm by introducing a compensation term to reduce
the tracking error under the time varying CSI.\end{abstract}
\maketitle
\IEEEpeerreviewmaketitle

\section{Introduction}

Saddle point problems arise in a number of wireless communication
applications such as resource allocations and competitive games. Resource
allocations problems can be formulated as a constrained maximization
of some utility functions. By constructing a Lagrangian function,
the constrained problem can be reformulated into an unconstrained
one and be solved by computing the saddle point of the Lagrangian
function, where we maximize over the primal variables and minimize
over the dual variables. Most remarkably, primal-dual gradient methods
have been widely used for computing the saddle points of general Lagrangian
functions, while they provide decentralized solutions in wireless
applications. The primal-dual gradient methods update the primal and
dual variables simultaneously by evaluating the gradient of both the
primal function and dual function at the same time. A classical study
of primal-dual algorithms has been done by Arrow, Hurwicz and Uzawa
in their seminal work \cite{Arrow1958}, where they constructed a
continuous-time algorithm for general concave-convex functions and
derived the asymptotic properties using Lyapunov-like methods. They
also provided a global stability result on a discrete-time algorithm
with constant step sizes. Recently, Feijer \emph{et al.} \cite{Feijier2009,Feijer:2010uq}
have studied the stability of these primal-dual algorithm dynamics
and extended the results to various network resource allocations problems.

In the above literature, when people discuss the convergence behavior
of the primal-dual gradient algorithm, all the problem specific parameters
are considered to be time invariant. Even in most other literatures
studying optimization problems in wireless communications, problem
dependent parameters are merely considered to be \emph{quasi-static},
meaning that these time varying parameters are assumed to be constant
for a sufficiently long time until the algorithms converge. However,
this assumption is unrealistic, especially in wireless communication
scenarios. For instance, the operating environment in terms of the
channel state information (CSI) may be changing frequently, such that
the optimization problem varies from time to time. On the other hand,
as \emph{explicit message passing} may be involved in the iterations,
the optimization algorithms cannot always converge fast enough to
catch up with time varying effects, \textcolor{black}{especially
for large scale problems. As is shown in our numerical example, convergence
errors would lead to performance loss in a wireless communication
system.} However, it is yet unknown whether the algorithms converge
or not when the quasi-static assumption is dropped, even for strictly
convex problems. Therefore, it is highly important to the study the
convergence behavior, or robustness, for primal-dual gradient algorithms
under time varying CSI.

However, towards this end there are a lot of technical challenges,
some of which are listed below.
\begin{itemize}
\item \textbf{How to quantify the performance penalty due to the time varying
parameters}
\item \textbf{How to evaluate the cost-performance tradeoff}
\item \textbf{How to enhance the algorithm}
\end{itemize}

These questions are highly nontrivial due to the stochastic nature
of wireless communication problems as well as the complexities of
the algorithms that solve them. There are some preliminary works studying
the effects caused by time varying parameters. For example, for resource
allocation problems, the authors in \cite{Kelly:1998fk} had introduced
stochastic perturbations to the algorithm to represent random loads
entering the network, and their analyses were based on linearization
on the system. The authors in \cite{Zhang:2007fk} studied a stochastic
\textcolor{black}{network utility maximization (NUM) problem} with
noisy feedback input to the algorithm. However, \cite{Kelly:1998fk}
and \cite{Zhang:2007fk} did not consider the CSI being time varying
and hence their problems have static equilibrium points. In \cite{Cheng:2010},
the authors studied the performance of gradient algorithms for a game
problem with moving equilibria, and the authors in \cite{Fan20061837,Alpcan:2003fk}
studied multi-cell CDMA interference games under time varying channels.
While these work \cite{Cheng:2010,Fan20061837,Alpcan:2003fk} provide
some preliminary results for solving games under time varying CSI,
their techniques cannot be applied to general saddle point problems.

In this paper, we shall investigate the convergence behavior of primal-dual
algorithms for solving general saddle point problems under time varying
CSI. We first define an equivalent virtual dynamic system, following
which we study the stability of such virtual dynamic system based
on the Lyapunov theory \cite{Khalil1996} and the LaSalle's invariance
principle argument \cite{Wen:2002bh}. We model the dynamics of the
time varying CSI as an auto-regressive system, and the stability results
are derived by constructing Lyapunov functions along the algorithm
trajectories. Based on the stability results, we derive the convergence
properties of primal-dual algorithms. We also quantify the average
tracking errors in terms of the average exogenous excitations induced
to the CSI dynamics. Furthermore, to enhance the tracking performance,
we propose a novel adaptive algorithm with distributive implementation
solutions. As an illustration, we also consider two wireless application
examples, based on which we show that the numerical results match
with our analysis.

The paper is organized as follows. In Section \ref{sec:System-Model},
we introduce a general saddle point problem as well as the primal-dual
algorithm with two application examples in wireless communication.
We also introduce the time varying CSI model, and the virtual dynamic
systems. Then in Section \ref{sec:Convegenece-Analysis-unconstrained}
and Section \ref{sec:Convergence-Analsyiss-Degraded}, we analyze
the convergence behavior of the primal-dual algorithm for the strongly
concave-convex saddle point problem and degraded saddle point problem,
respectively. In Section \ref{sec:Adaptive-Proposed-alg}, we propose
a novel adaptive algorithm to reduce the tracking error in time varying
CSI. Section \ref{sec:Results-and-Discussions} gives the simulation
results and discussions. Finally, we summarize our main results in
Section \ref{sec:Conclusions}.

\emph{Notations: }Matrices and vectors are denoted by capitalized
and lowercase boldface letters, respectively. $A^{T}(\mathbf{a}^{T})$
denotes the transpose of matrices (vectors) $A$ ($\mathbf{a}$),
and $A^{H}$ denotes the complex conjugate transpose. $\mathbb{R}$
and $\mathbb{C}$ denote the set of real numbers and complex numbers,
respectively. $\mathbf{a}\succeq\mathbf{b}$ denotes the entry-by-entry
comparison and $A\succeq B$ means that $A-B$ is positive semidefinite.
Unless specified in the context, $\|\centerdot\|$ denotes the Euclidean
norm.

\section{System Model and Virtual Dynamic Systems\label{sec:System-Model}}

In this section, we shall first introduce a general saddle point problem
setup and illustrate with a few examples in the context of wireless
communications. We shall then define the time varying CSI model and
discuss its impact on the convergence of saddle point problems. Finally,
we shall introduce the notion of \emph{virtual dynamic system} and
establish the equivalence between convergence behavior of the saddle
point algorithm and the stability of the\emph{ virtual dynamic system}.

\subsection{General Saddle Point Problem\label{sub:SysModel-General-Saddle-Point}}

We consider a min-max optimization problem 
\begin{equation}
\min_{\mathbf{\lambda}\in\mathbb{R}_{+}^{m}}\max_{\mathbf{x}\in\mathbb{R}^{n}}\mathcal{L}(\mathbf{x},\,\mathbf{\mathbf{\lambda}};\,\mathbf{h})\label{eqn:min-max}
\end{equation}
where the function $\mathcal{L}:\,\mathbb{R}^{n}\times\mathbb{R}_{+}^{m}\mapsto\mathbb{R}$
has two vector variables. Moreover, $\mathcal{L}(\mathbf{x},\,\mathbf{\lambda};\,\mathbf{h})$
is strongly concave in $\mathbf{x}\in\mathbb{R}^{n}$ and convex in
$\mathbf{\lambda}\in\mathbb{R}_{+}^{m}$. $\mathbf{h}\in\mathcal{H}\subseteq\mathbb{R}^{q}$
is a vector parameter that arises from specific optimization problems.
In the context of wireless communication optimizations, the problem
parameter $\mathbf{h}$ can be the CSI. Under the above convexity
assumption, the min-max problem has a unique optimal solution $(\mathbf{x}^{*},\,\lambda^{*})$.
Note that since the min-max problem in \eqref{eqn:min-max} is parameterized
by $\mathbf{h}$, the optimal solution $(\mathbf{x}^{*},\,\lambda^{*})$
can also be represented as a mapping from $ $$\mathbf{h}\in\mathcal{H}$
to $(\mathbf{x}^{*},\,\lambda^{*})$ 
\[
\mathbf{x}^{*}(\mathbf{h})=\psi_{x}(\mathbf{h}),\;\mathbf{\lambda}^{*}(\mathbf{h})=\psi_{\lambda}(\mathbf{h})\quad\forall\mathbf{h}\in\mathcal{H}
\]
where $\psi_{x}:\mathcal{H}\mapsto\mathbb{R}^{n}$ and $\psi_{\lambda}:\mathcal{H}\mapsto\mathbb{R}_{+}^{m}$
are $\mathcal{C}^{1}$ functions.

It is known that solving the above min-max optimization problem \eqref{eqn:min-max}
is equivalent to computing the saddle point of $\mathcal{L}(\mathbf{x},\,\mathbf{\lambda};\,\mathbf{h})$
\cite{Boyd:2004kx}. For a given $\mathbf{h}\in\mathcal{H}$, a saddle
point $(\mathbf{x}^{*},\,\mathbf{\lambda}^{*})$ of $\mathcal{L}(\mathbf{x},\,\mathbf{\lambda};\,\mathbf{h})$
over the set $\mathbb{R}^{n}\times\mathbb{R}_{+}^{m}$ is defined
to be a vector that satisfies 
\[
\mathcal{L}(\mathbf{x},\,\mathbf{\lambda}^{*};\,\mathbf{h})\leq\mathcal{L}(\mathbf{x}^{*},\,\mathbf{\lambda}^{*};\,\mathbf{h})\leq\mathcal{L}(\mathbf{x}^{*},\,\mathbf{\lambda};\,\mathbf{h}),\quad\forall\mathbf{x}\in\mathbb{R}^{n},\,\mathbf{\lambda}\in\mathbb{R}_{+}^{m}.
\]
It is also known that when $\mathcal{L}$ is strongly concave in $\mathbf{x}$
and convex in $\mathbf{\lambda}$, the saddle point is unique \cite{Boyd:2004kx},
meaning that the optimal solution to \eqref{eqn:min-max} is unique.

The classical primal-dual gradient algorithm dynamics%
\footnote{In this paper, we assume the step size of iterations are sufficiently
small so that the primal-dual iterations can be represented by the
continuous time dynamics \cite{Arrow1958,Feijer:2010uq}.%
} for solving \eqref{eqn:min-max} are given in the following: 
\begin{eqnarray}
\dot{\mathbf{x}} & = & \frac{d\mathbf{x}}{dt}=\kappa\left[\frac{\partial}{\partial\mathbf{x}}\mathcal{L}(\mathbf{x},\,\mathbf{\mathbf{\lambda}};\,\mathbf{h}(t))\right]^{T}\label{eqn:alg_Arrow-gradient-h(t)-1}\\
\dot{\mathbf{\lambda}} & = & \frac{d\mathbf{\lambda}}{dt}=\kappa\left[\left(-\frac{\partial}{\partial\mathbf{\lambda}}\mathcal{L}(\mathbf{x},\,\mathbf{\mathbf{\lambda}};\,\mathbf{h}(t))\right)^{T}\right]_{\lambda}^{+}\label{eqn:alg_Arrow-gradient-h(t)-2}
\end{eqnarray}
for some step size $\kappa>0$. The projection $\left[\centerdot\right]_{\lambda}^{+}$
is to restrict $\mathbf{\lambda}$ in the nonnegative domain $\mathbb{R}_{+}^{m}$.
For scalars $u_{i}$ and $\lambda_{i}$, the projection is defined
to be $[u_{i}]_{\lambda_{i}}^{+}:=u_{i}$ if $u_{i}>0$ or $\lambda_{i}>0$,
and $[u_{i}]_{\lambda_{i}}^{+}:=0$ otherwise. For the vector case,
the projection is defined entry-wide. This algorithm was first introduced
by Arrow \emph{et al.} \cite{Arrow1958} and has been recently studied
and applied to various resource allocations applications \cite{Feijer:2010uq}.
The asymptotic behavior has been analyzed by studying the Lyapunov
function of the system state and applying LaSalle's \emph{invariant
principle} argument \cite{Khalil1996,Wen:2002bh}. 

In all these works, the convergence results were established based
on the assumption that the wireless channel state $\mathbf{h}$ stays
time invariant before the algorithm converges. In practice, the iterations
in \eqref{eqn:alg_Arrow-gradient-h(t)-1}-\eqref{eqn:alg_Arrow-gradient-h(t)-2}
may involve explicit message passing among nodes in a wireless network,
and hence it is quite unlikely for the $\mathbf{h}$ to be invariant
during the iterations. When the channel state \textcolor{black}{$\mathbf{h}(t)$}
is time varying, all the existing convergence results \cite{Arrow1958,Feijer:2010uq}
fail to apply. In this paper, we shall address the situation where
the CSI \textcolor{black}{$\mathbf{h}(t)$} changes in a similar
time scale as the primal-dual algorithm in \eqref{eqn:alg_Arrow-gradient-h(t)-1}-\eqref{eqn:alg_Arrow-gradient-h(t)-2},
which reflects a more realistic situation in practice.

\subsection{Examples of Saddle Point Problems in Wireless Communications\label{sub:System Model-Illustration-of-Saddle}}

In this subsection, we shall illustrate various important applications
of saddle point problems in the context of wireless communications.

\subsubsection{Application 1: Transmission and Jamming Strategy Optimization in
MIMO Channels\label{subsub:App_Transmission_Jamming}}

\begin{figure}
\centering{}\includegraphics[scale=0.5]{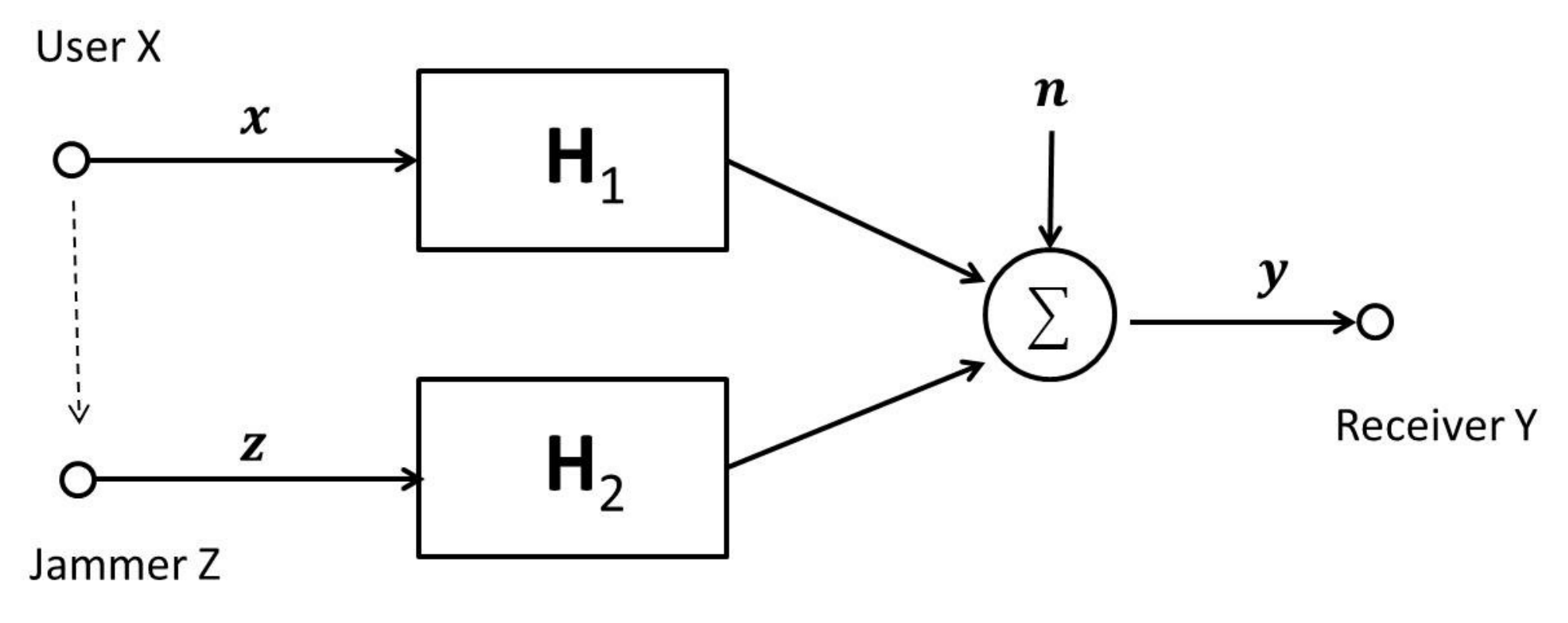}\caption{\label{fig:Jammer-system-model}A system model for a point-to-point
MIMO with a jammer. The user X transmits signal to Y through the channel
$\mathbf{H}_{1}$ and a jammer Z transmits a jamming signal to Y through
the channel $\mathbf{H}_{2}$. }
\end{figure}

Consider a point-to-point MIMO system with a jammer as shown in Fig.
\ref{fig:Jammer-system-model}. The transmitter and the receiver have
$N$ antennas. The user X transmits the desired signal $\mathbf{x}\thicksim\mathcal{CN}(0,\mathbf{Q})$
to the receiver Y, while the jammer Z transmits a jamming signal $\mathbf{z}\thicksim\mathcal{CN}(0,\mathbf{Z})$
to interfere Y. The received signal at Y is given by $\mathbf{y}=\mathbf{H}_{1}\mathbf{x}+\mathbf{H}_{2}\mathbf{z}+\mathbf{n}$,
where $\mathbf{n}\thicksim\mathcal{CN}(0,\sigma_{n}^{2}\mathbf{I})$
is the additive Gaussian noise, and the mutual information between
X and Y is given by \cite{Song:2002fk} 
\begin{equation}
C(\mathbf{Q},\,\mathbf{Z})=\log\det\left(\mathbf{I}+\left(\sigma_{n}^{2}\mathbf{I}+\mathbf{H}_{2}\mathbf{Z}\mathbf{H}_{2}^{H}\right)^{-1}\mathbf{H}_{1}\mathbf{Q}\mathbf{H}_{1}^{H}\right).\label{eqn:exp2-capacity}
\end{equation}
Assume that both the user and jammer have perfect transmit CSI (CSIT),
and they know each other's transmission covariance. Thus the user's
strategy is to maximize the mutual information given in \eqref{eqn:exp2-capacity}
based on the CSIT and the knowledge of the jammer's transmission covariance,
while the the jammer's strategy is to minimize that mutual information.
Their strategies result in finding the saddle point of the following
saddle point problem \cite{Jorswieck2005,Kashyap2004}.
\begin{problem}
[Transmission and Jamming Strategy Optimization]\label{pro:Transmission-and-Jamming}
\begin{equation}
\min_{tr(\mathbf{Z})\leq P_{\mathcal{J}}}\max_{tr(\mathbf{Q})\leq P_{T}}C(\mathbf{Q},\,\mathbf{Z};\,\mathbf{h})=\log\det\left(\mathbf{I}+\left(\sigma_{n}^{2}\mathbf{I}+\mathbf{H}_{2}\mathbf{Z}\mathbf{H}_{2}^{H}\right)^{-1}\mathbf{H}_{1}\mathbf{Q}\mathbf{H}_{1}^{H}\right).\label{eqn:exp2-saddle point prob}
\end{equation}
where $\mathbf{h}=\mbox{vec}\left([\mathbf{H}_{1}\,\mathbf{H}_{2}]\right)$,
$P_{T}$ and $P_{\mathcal{J}}$ are the total power constraints for
the user and jammer, respectively.
\end{problem}

Note that the saddle point problem in \eqref{eqn:exp2-saddle point prob}
is strongly concave in $\mathbf{Q}$ and strongly convex in $\mathbf{Z}$
\cite{Kashyap2004}, and hence there exists a unique saddle point,
which can be computed by a primal-dual iterative algorithm \eqref{eqn:alg_Arrow-gradient-h(t)-1}-\eqref{eqn:alg_Arrow-gradient-h(t)-2}.
When the CSI $\mathbf{h}(t)$ is time invariant, the transmission
policies for both the user and jammer will converge to the saddle
point of \eqref{eqn:exp2-saddle point prob}, and a Nash Equilibrium
is achieved. However, when the CSI $\mathbf{h}(t)$ is time varying
in a similar timescale as the primal-dual iterations, the convergence
is not guaranteed. Hence, it is interesting to study the convergence
behavior of the algorithm iterations under time-varying CSI.

\subsubsection{Application 2: Network Utility Maximization of Wireless Ad Hoc Network\label{subsub:App_NUM}}

\begin{figure}
\begin{centering}
\includegraphics[scale=0.35]{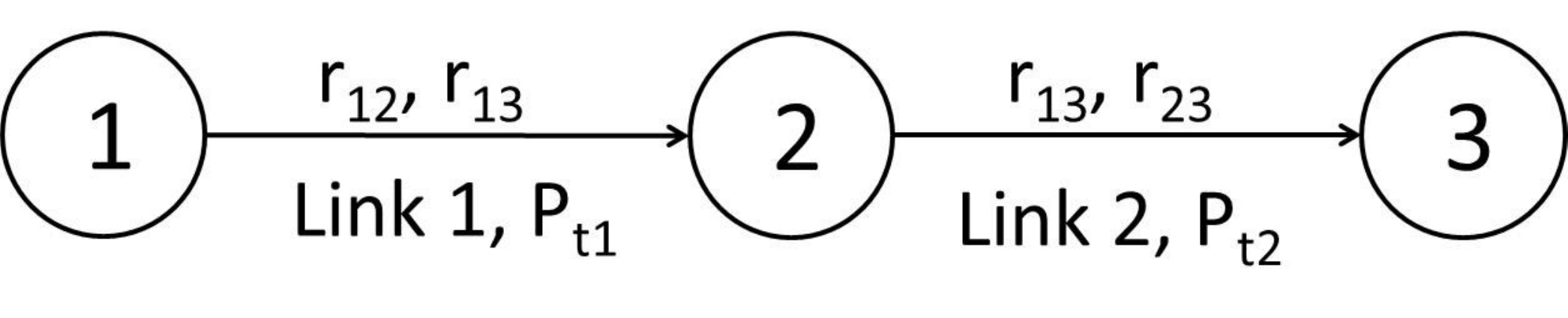} 
\par\end{centering}

\caption{\label{fig:Multihop wireless network}A specific example of the multihop
wireless network with $3$ nodes and $2$ links, where the collection
of traffic flows is given by $\mathcal{C}=\{(1,2),\,(1,3),\,(2,3)\}$,
and the sets of links are $L(1,2)=\{1\}$, $L(1,3)=\{1,2\}$ and $L(2,3)=\{2\}$.
$P_{t1}$ and $P_{t2}$ denote the total power allocated to link 1
and link 2 respectively. }
\end{figure}

In a resource allocation problem, the communication network is modeled
as a set of source nodes transmitting traffic through a set of links.
Each source node has its own concave utility function and is allocated
a portion of resources. The optimization problem is to maximize the
total network utility under a set of constraints on the total network
resources. The problem is formulated as follows \cite{Kelly:1998fk,Palomar:2007vn}.
\begin{problem}
[Network Utility Maximization]\label{pro:Network-Utility-Maximization}

\begin{eqnarray}
\max_{\mathbf{r}\succeq0,\,\mathbf{p}\succeq0} & \sum_{(s,\, d)\in\mathcal{C}}U_{sd}(r_{sd})\label{eqn:pro1-formulation}\\
\text{subject to} & \sum_{(s,\, d):\, l\in L(s,\, d)}r_{sd}\leq c_{l}(p_{l};\,\mathbf{h}) & \forall l\nonumber \\
 & \mathbf{p}\in\mathcal{P}\nonumber 
\end{eqnarray}
where the ordered pair $(s,\, d)$ denotes the traffic, \textcolor{black}{each
of which has a fixed route,} that initiates from source node $s$
and is delivered to destination node $d$, $r_{sd}$ denotes the data
rate for the traffic, $\mathbf{p}=[p_{1}\,\dots\, p_{L}]^{T}$ is
a vector of power allocated to $L$ links, $U_{sd}$ is the utility
function to evaluate traffic rate $r_{sd}$ for source $s$, and $c_{l}(\centerdot)$
is a link capacity function of the transmit power $p_{l}$ and channel
gain (i.e. CSI) $\mathbf{h}$. $\mathcal{C}$ denotes the collection
of all traffic flows, $L(s,\, d)$ denotes the set of links that traffic
$(s,\, d)$ goes through, and $\mathcal{P}$ denotes the set of feasible
power vectors. We assume both $U_{sd}(\centerdot)$ and $c_{l}(\centerdot)$
are strongly concave and twice differentiable. Fig. \ref{fig:Multihop wireless network}
illustrates an example NUM problem with $3$ nodes, $2$ links, and
$|\mathcal{C}|=3$ traffic flows. For example, the collection of traffic
flows is given by $\mathcal{C}=\{(1,2),\,(1,3),\,(2,3)\}$, and the
sets of links are $L(1,2)=\{1\}$, $L(1,3)=\{1,2\}$ and $L(2,3)=\{2\}$.
\end{problem}

For notation convenience, we denote $\mathbf{x}=[\mathbf{r}\,;\,\mathbf{p}]$
as a vector of the primal variables. The Lagrangian of \eqref{eqn:pro1-formulation}
can be written as 
\begin{eqnarray}
\mathcal{L}(\mathbf{r},\mathbf{p},\,\mathbf{\lambda};\,\mathbf{h}) & = & \sum_{s\in\mathcal{S}}U(r_{s})-\sum_{l}\lambda_{l}\left(\sum_{s:\, l\in L(s)}r_{s}-c_{l}(p_{l};\,\mathbf{h})\right)\label{eqn:pro1-Lagrangian}
\end{eqnarray}
\textcolor{black}{subjected to the power constraint $\mathbf{p}\in\mathcal{P}$.
}From the Lagrangian theory \cite{Boyd:2004kx}, solving the optimization
problem \eqref{eqn:pro1-formulation} is equivalent to finding the
saddle point of the Lagrangian function \eqref{eqn:pro1-Lagrangian},\textcolor{black}{
\begin{equation}
\min_{\lambda\succeq0}\max_{\mathbf{r}\succeq0;\mathbf{p}\in\mathcal{P}}\mathcal{L}(\mathbf{r},\mathbf{p},\,\lambda;\,\mathbf{h})\label{eqn:pro1-saddle point prob}
\end{equation}
}which is a special instance of \eqref{eqn:min-max}, and the primal-dual
algorithm \eqref{eqn:alg_Arrow-gradient-h(t)-1}-\eqref{eqn:alg_Arrow-gradient-h(t)-2}
can be applied to solve this problem. 

Note that as solving the dual problem involves explicit message passing
among nodes, the convergence rate of the algorithm is critical to
the solution quality of the NUM problem. Since it may not be justified
in practice that the channel state $\mathbf{h}(t)$ is time invariant
during the algorithm convergence, it is very important to study the
convergence behavior of the primal-dual iteration in \eqref{eqn:alg_Arrow-gradient-h(t)-1}-\eqref{eqn:alg_Arrow-gradient-h(t)-2}
when the CSI is time varying.

\subsection{Time Varying CSI Model}

In this section, we first introduce a dynamic model for the time-varying
CSI $\mathbf{h}(t)$. Specifically, the CSI $\mathbf{h}(t)$ is modeled
as a solution to the following dynamic equation 
\begin{equation}
\dot{\mathbf{h}}(t)=\frac{d\mathbf{h}(t)}{dt}=A(\mathbf{h}(t)-\bar{\mathbf{h}})+u(t)\label{sys:h-model-general}
\end{equation}
where $A$ is a real symmetric negative definite matrix, $u(t)$ is
a vector valued \textcolor{black}{complex Gaussian process with uncorrelated
real and imaginary components,} and $\bar{\mathbf{h}}$ is a constant
vector, corresponding to the line-of-sight (LOS) component in the
channel model. The dynamic system can be viewed as an external disturbance
$u(t)$ being applied to an autonomous%
\footnote{An autonomous system is a system that can be described by an ordinary
differential equation $\dot{x}=f(x)$ where $f$ does not explicitly
depend on $t$ \cite{Khalil1996}.%
} linear system $\dot{\mathbf{h}}=A(\mathbf{h}(t)-\bar{\mathbf{h}})$.
Note that the dynamic model \eqref{sys:h-model-general} resembles
an AR(1) process \textcolor{black}{and $|\mathbf{h}(t)|$ has a stationary
Rician distribution. As is summarized in \cite{Baddour01}, the AR
models have been successfully used to model, predict and simulate
fading channel dynamics (e.g. in \cite{Tsatsanis00,Baddour01,Wu00}),
the autocorrelation function of which was also observed and justified
in \cite{wu96} to be close to that of a Rayleigh fading process.}

Note that the dynamic system model of $\mathbf{h}$ in \eqref{sys:h-model-general}
is stable%
\footnote{The concept of stability is formally introduced in Definition \ref{def:Stability}
in the following subsection.%
} \cite{Khalil1996}. Firstly, if $u(t)=0$, the dynamic system \eqref{sys:h-model-general}
exponentially converges to the unique equilibrium point $\mathbf{h}=\bar{\mathbf{h}}$.
Secondly, when $u(t)\neq0$, it is also stable provided that $\overline{\|u(t)\|^{2}}=\lim_{T\to\infty}\frac{1}{T}\int_{0}^{T}\|u(t)\|^{2}dt<\infty$.
This can be verified by constructing a Lyapunov function $L(\mathbf{h})=(\mathbf{h}-\bar{\mathbf{h}})^{T}(\mathbf{h}-\bar{\mathbf{h}})/2$
on the state $\mathbf{h}$ and applying the LaSalle's \emph{invariant
principle} argument \cite{Wen:2002bh,Khalil1996}. The stability property
in the CSI model $\mathbf{h}(t)$ is important because it is a necessary
condition for the convergence of the primal-dual algorithm \eqref{eqn:alg_Arrow-gradient-h(t)-1}-\eqref{eqn:alg_Arrow-gradient-h(t)-2}.

\subsection{Virtual Dynamic Systems\label{sub:Virtual-Dynamic-Systems}}

In this section, we shall first define a \emph{virtual dynamic system}.
We shall illustrate that studying the convergence behavior of the
primal-dual iterative algorithm for the saddle point problem in \eqref{eqn:alg_Arrow-gradient-h(t)-1}-\eqref{eqn:alg_Arrow-gradient-h(t)-2}
is the same as studying the stability behavior of the virtual dynamic
system. As a result, the virtual dynamic system forms a bridge between
the convergence analysis of iterative algorithms and nonlinear control
theory. 
\begin{lyxDefQED}
[Virtual Dynamic System]\label{def:Virtual-Dynamic-System} Let
$\mathbf{\widetilde{x}}=(\mathbf{x},\,\lambda)$ be the joint state
of the primal and dual variables in algorithm \eqref{eqn:alg_Arrow-gradient-h(t)-1}-\eqref{eqn:alg_Arrow-gradient-h(t)-2}
and let $\widetilde{f}:\,\mathbb{R}^{n}\times\mathbb{R}_{+}^{m}\mapsto\mathbb{R}^{n+m}$
be the vector valued function 
\begin{equation}
\widetilde{f}(\widetilde{\mathbf{x}};\,\mathbf{h}(t))\triangleq\left[\begin{array}{c}
\kappa\left(\frac{\partial}{\partial\mathbf{x}}\mathcal{L}(\mathbf{x},\,\mathbf{\mathbf{\lambda}};\,\mathbf{h}(t))\right)^{T}\\
\kappa\left[\left(-\frac{\partial}{\partial\mathbf{\lambda}}\mathcal{L}(\mathbf{x},\,\mathbf{\mathbf{\lambda}};\,\mathbf{h}(t))\right)^{T}\right]_{\lambda}^{+}
\end{array}\right].\label{eqn:vector field for a virtual dynamic sys}
\end{equation}
The state $\widetilde{\mathbf{x}}$ of the \emph{virtual dynamic system}
satisfies the following dynamic equation 
\begin{equation}
\widetilde{\mathcal{X}}:\quad\dot{\widetilde{\mathbf{x}}}=\frac{d}{dt}\mathbf{\widetilde{x}}(t)=\widetilde{f}(\mathbf{\widetilde{x}};\,\mathbf{h}(t)).\label{eqn:def-virtual dynamic system}
\end{equation}

\end{lyxDefQED}
\begin{lyxDefQED}
[Equilibrium Point] \label{def:Equilibrium-Point} $\widetilde{\mathbf{x}}^{*}$
is called an equilibrium point of the vector field $\dot{\widetilde{\mathbf{x}}}=\widetilde{f}(\mathbf{\widetilde{x}};\,\mathbf{h}(t))$,
if $\widetilde{f}(\mathbf{\widetilde{x}}^{*};\,\mathbf{h}(t))=0$.
\end{lyxDefQED}

Note that since the vector field $\dot{\widetilde{\mathbf{x}}}=\widetilde{f}(\mathbf{\widetilde{x}};\,\mathbf{h})$
is parameterized by the channel state $\mathbf{h}(t)$, the equilibrium
point $\widetilde{\mathbf{x}}^{*}(\mathbf{h}(t))$ is expressed as
a function of $\mathbf{h}(t)$. From \eqref{eqn:def-virtual dynamic system}
and Definition \ref{def:Equilibrium-Point}, we have the following
result establishing the connection between the virtual dynamic system
\eqref{eqn:def-virtual dynamic system} and the primal-dual algorithm
\eqref{eqn:alg_Arrow-gradient-h(t)-1}-\eqref{eqn:alg_Arrow-gradient-h(t)-2}
for solving saddle point problems. 
\begin{lyxThmQED}
[Connections between the Virtual Dynamic System and the Saddle Point
Problem]\label{thm:Connections-between-Virtual-Saddle point} The
\emph{equilibrium point} $\widetilde{\mathbf{x}}^{*}(\mathbf{h}(t))$
of the virtual dynamic system \eqref{eqn:def-virtual dynamic system}
is identical to the saddle point $\left[\mathbf{x}^{*}(\mathbf{h}(t)),\,\mathbf{\lambda^{*}(}\mathbf{h}(t))\right]$
of the saddle point problem \eqref{eqn:min-max}. Furthermore, the
algorithm trajectories of the primal-dual algorithm in \eqref{eqn:alg_Arrow-gradient-h(t)-1}-\eqref{eqn:alg_Arrow-gradient-h(t)-2}
are the same as those of the virtual dynamic system \eqref{eqn:def-virtual dynamic system}.\end{lyxThmQED}
\begin{proof}
Please refer to Appendix \ref{sec:App_Proof-of-Theorem_connection}
for the proof. 
\end{proof}

\begin{figure}
\begin{centering}
\includegraphics[scale=0.55]{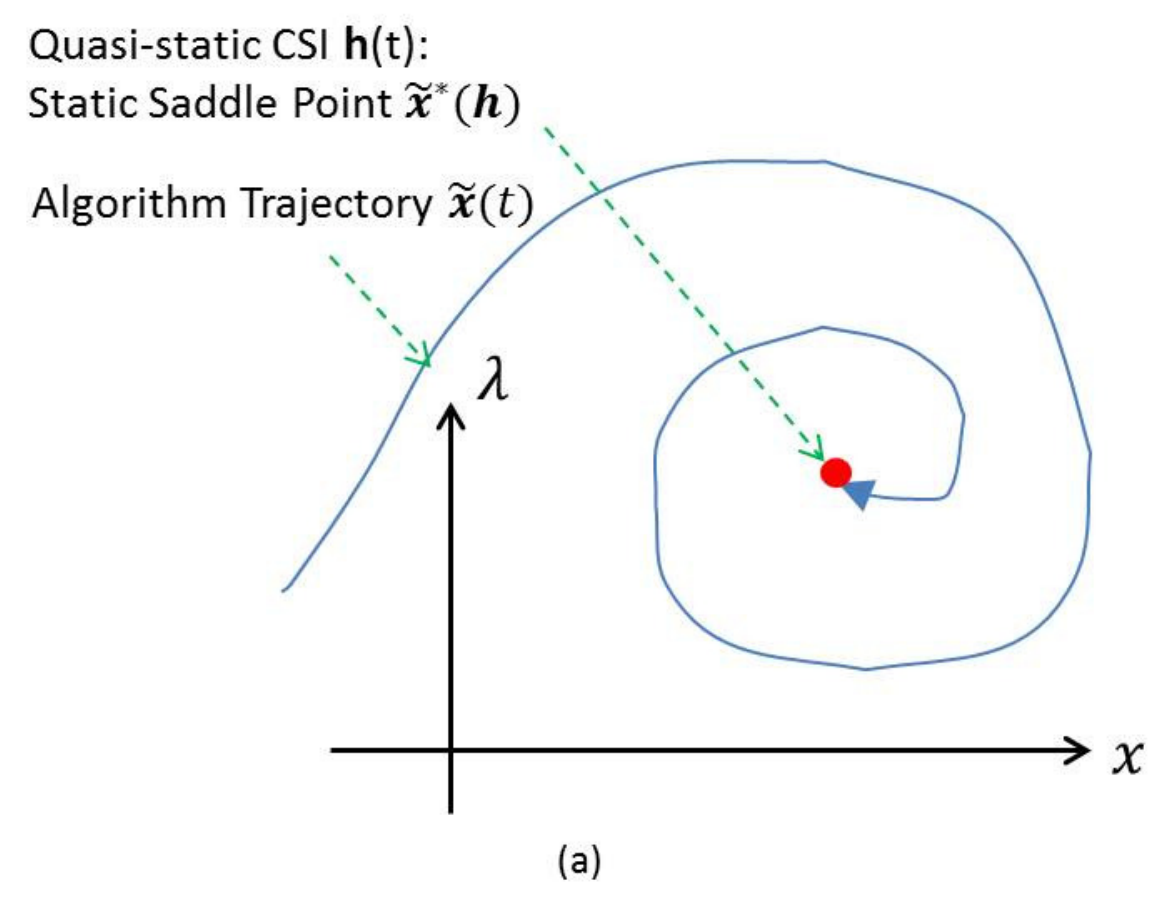}\includegraphics[scale=0.55]{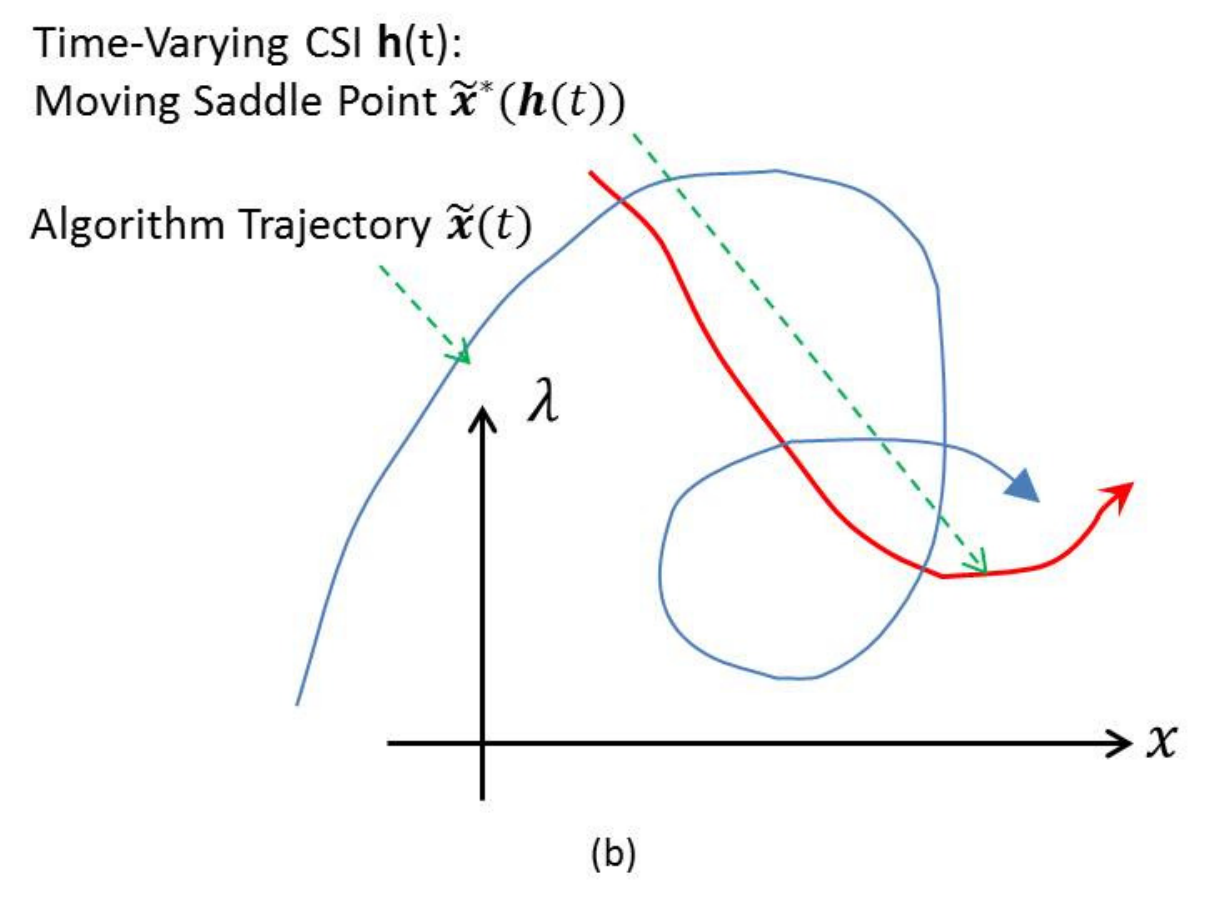}
\par\end{centering}

\caption{\label{fig:moving-equilibrium}An illustration of the convergence
behavior of the primal-dual algorithms. In (a), the equilibrium point
in the equivalent virtual dynamic system is asymptotically stable
when the CSI $\mathbf{h}(t)$ is quasi-static \cite{Arrow1958}. However,
in (b), when the CSI $\mathbf{h}(t)$ is time-varying, the equilibrium
point $\widetilde{\mathbf{x}}^{*}(t)$ becomes also time-varying and
the convergence is not guaranteed.}

\end{figure}

As a result of Theorem \ref{thm:Connections-between-Virtual-Saddle point},
the convergence behavior of the primal-dual algorithm \eqref{eqn:alg_Arrow-gradient-h(t)-1}-\eqref{eqn:alg_Arrow-gradient-h(t)-2}
can be visualized by the stability of the equilibrium point in the
virtual dynamic system \eqref{eqn:def-virtual dynamic system}. For
instance, the algorithm in \eqref{eqn:alg_Arrow-gradient-h(t)-1}-\eqref{eqn:alg_Arrow-gradient-h(t)-2}
converges to the saddle point if and only if the equilibrium point
of the virtual system \eqref{eqn:def-virtual dynamic system} is asymptotically
stable. When the channel state $\mathbf{h}$ is time invariant, the
equilibrium point $\widetilde{\mathbf{x}}^{*}(\mathbf{h})$ is stationary
and the algorithm trajectory of the primal-dual iterations can be
visualized to converge to the stationary equilibrium point. On the
other hand, when the channel state $\mathbf{h}(t)$ is time varying
(with comparable time scale as the algorithm trajectory), the equilibrium
point $\widetilde{\mathbf{x}}^{*}(\mathbf{h}(t))$ will be time varying
and the convergence of the algorithm cannot be guaranteed. This is
illustrated in Fig. \ref{fig:moving-equilibrium}. Due to this association,
we shall focus on studying the stability of the \emph{moving equilibrium}
$\widetilde{\mathbf{x}}^{*}(\mathbf{h}(t))$ of the virtual system
\eqref{eqn:def-virtual dynamic system} in the rest of the paper. 

Due to the time varying CSI, the equilibrium point $\widetilde{\mathbf{x}}^{*}(\mathbf{h}(t))$
is non-stationary. For mathematical convenience, we consider a change
of state variable. Define $\mathbf{\widetilde{x}}_{e}(t)=\mathbf{\widetilde{x}}(t)-\widetilde{\mathbf{x}}^{*}(t)$.
Note that $\widetilde{\mathbf{x}}_{e}(t)$ represents the instantaneous
error%
\footnote{Using Theorem \ref{thm:Connections-between-Virtual-Saddle point},
$\widetilde{\mathbf{x}}_{e}(t)$ also represents the error between
the current solution $(\mathbf{x}(t),\,\lambda(t))$ driven by the
iterative algorithm in \eqref{eqn:alg_Arrow-gradient-h(t)-1}-\eqref{eqn:alg_Arrow-gradient-h(t)-2}
and the instantaneous saddle point $(\mathbf{x}^{*}(t),\,\lambda^{*}(t))$. %
}. Let $\psi(\mathbf{h}):=(\mathbf{x}^{*}(\mathbf{h}),\,\lambda^{*}(\mathbf{h}))=\widetilde{\mathbf{x}}^{*}(\mathbf{h})$
be a mapping from the CSI $\mathbf{h}\in\mathcal{H}$ to the equilibrium
point of the virtual dynamic system \eqref{eqn:def-virtual dynamic system}.
As $\dot{\widetilde{\mathbf{x}}}_{e}=\dot{\widetilde{\mathbf{x}}}-\dot{\widetilde{\mathbf{x}}}^{*}$,
The error process $\widetilde{\mathbf{x}}_{e}(t)$ satisfies the dynamic
equation of the \emph{error dynamic system }defined below.
\begin{lyxDefQED}
[Error Dynamic System]\label{def:Tracking-Error-Dynamic-sys} The
error dynamic system for the virtual system $\dot{\widetilde{\mathbf{x}}}=\widetilde{f}(\mathbf{\widetilde{x}};\,\mathbf{h}(t))$
is given by
\begin{equation}
\widetilde{\mathcal{X}}_{e}:\quad\dot{\widetilde{\mathbf{x}}}_{e}=\widetilde{f}_{e}(\widetilde{\mathbf{x}}_{e};\,\mathbf{h}(t))-\varphi(\mathbf{h})\dot{\mathbf{h}}(t)\label{sys:error-dynamic-system}
\end{equation}
where $\widetilde{f}_{e}(\mathbf{\widetilde{x}}_{e};\,\mathbf{h}(t)):=\widetilde{f}\left(\widetilde{\mathbf{x}}_{e}+\psi(\mathbf{h});\,\mathbf{h}(t)\right)$
and $\varphi(\mathbf{h}):=\frac{\partial}{\partial\mathbf{h}}\psi(\mathbf{h})$
.
\end{lyxDefQED}

Since the error state $\widetilde{\mathbf{x}}_{e}(t)$ and the virtual
system state $\widetilde{\mathbf{x}}(t)$ are related simply by a
linear transformation, studying the stability of $\widetilde{\mathcal{X}}$
is equivalent to studying the stability of $\widetilde{\mathcal{X}}_{e}$.
Finally, we summarize the formal definition of as follows \cite{Khalil1996}.
\begin{lyxDefQED}
[Exponential Stability of the Virtual Dynamic System]\label{def:Stability}The
equilibrium point $\widetilde{\mathbf{x}}_{e}^{*}=0$ of \eqref{sys:error-dynamic-system}
is exponentially stable if there exist some positive constants $c$,
$k$, and $\lambda$ such that 
\begin{equation}
\|\widetilde{\mathbf{x}}_{e}(t)\|\leq k\|\widetilde{\mathbf{x}}_{e}(t_{0})\|e^{-\lambda(t-t_{0})},\;\forall\|\widetilde{\mathbf{x}}_{e}(t_{0})\|<c.\label{eqn:def_exp-stable}
\end{equation}
It is globally exponentially stable if \eqref{eqn:def_exp-stable}
is satisfied for any initial state $\widetilde{\mathbf{x}}_{e}(t_{0})$.
\end{lyxDefQED}

Using Theorem \ref{thm:Connections-between-Virtual-Saddle point},
we have global asymptotic stability of the virtual dynamic system
corresponding to the global asymptotic convergence of the primal-dual
algorithm in the original saddle point problem and vice versa.

\section{Convergence Analysis of Strongly Concave-Convex Saddle Point Problems
under Time Varying CSI\label{sec:Convegenece-Analysis-unconstrained}}

In this section, we shall focus on the convergence analysis of strongly
concave-convex saddle point problems when the CSI is time varying.
In this case, for the saddle point function $\mathcal{L}$, the Hessian
$\nabla_{x}^{2}\mathcal{L}\preceq-M_{x}\mathbf{I}$ for $M_{x}>0$
is negative definite and $\nabla_{\lambda}^{2}\mathcal{L}\succeq M_{\lambda}\mathbf{I}$
for $M_{\lambda}>0$ is positive definite. We shall first establish
the key stability results of the virtual dynamic system under a virtual
exogenous input. Based on that, we shall discuss the impact of time
varying CSI to the convergence of the saddle point problem, followed
by some examples.

\subsection{Stability Analysis of Virtual Dynamic System}

We shall study the stability of the virtual dynamic system $\widetilde{\mathcal{X}}_{e}$
in \eqref{sys:error-dynamic-system} associated with the primal-dual
iterative algorithm of a strongly concave-convex saddle point problem
in \eqref{eqn:alg_Arrow-gradient-h(t)-1}-\eqref{eqn:alg_Arrow-gradient-h(t)-2}.
We first consider the case when the CSI is time invariant ($\dot{\mathbf{h}}=0$).
The following lemma summarizes the key results. 
\begin{lyxLemQED}
[Stability of $\widetilde{\mathcal{X}}_{e}$ with Time Invariant
CSI]\label{lem:stability of x_e with time-invariant CSI}When the
CSI $\mathbf{h}(t)$ is time invariant, the equilibrium point $\widetilde{\mathbf{x}}_{e}^{*}=0$
of the virtual dynamic system $\widetilde{\mathcal{X}}_{e}$ in \eqref{sys:error-dynamic-system}
is globally exponentially stable for all $\mathbf{h}\in\mathcal{H}$. \end{lyxLemQED}
\begin{proof}
Please refer to Appendix \ref{sec:App_Proof-of-Lemma_stability_x_e}
for the proof.
\end{proof}

The above convergence result is not surprising since the CSI is time
invariant and similar results have been established in \cite{Arrow1958,Feijer:2010uq,Boyd:2004kx}
using different approaches. In the following, we shall extend the
stability result to the system with time varying CSI. 

As an intermediate, we first consider a virtual dynamic system $\widetilde{\mathcal{X}}_{e}$
in \eqref{sys:error-dynamic-system} with a \emph{quasi-time varying
}CSI\emph{ }$\mathbf{h}(t)$. Specifically, $\mathbf{h}(t)$ is time
varying according to the CSI model in \eqref{sys:h-model-general}
with $u(t)=0$, for all $t$. As a result, the CSI $\mathbf{h}(t)$
varies only during some transient and will converge to $\bar{\mathbf{h}}$
as $t\to\infty$. Define a joint system state as $\widetilde{\mathbf{z}}_{e}=(\widetilde{\mathbf{x}}_{e},\,\mathbf{h}_{e})$,
where $\mathbf{h}_{e}=\mathbf{h}-\bar{\mathbf{h}}$, the virtual dynamic
system of state $\widetilde{\mathbf{z}}_{e}$ can be specified as

\begin{equation}
\widetilde{\mathcal{Z}}_{e}:\quad\dot{\mathbf{\widetilde{z}}}_{e}=\widetilde{\mathcal{Z}}(\widetilde{\mathbf{z}}_{e}):=\left[\begin{array}{c}
\widetilde{f}_{e}(\mathbf{\widetilde{x}}_{e};\,\mathbf{h}_{e})+\varphi(\mathbf{h}_{e}+\bar{\mathbf{h}})A\mathbf{h}_{e}\\
A\mathbf{h}_{e}
\end{array}\right].\label{sys:Z(x_e,h)}
\end{equation}
\textcolor{black}{where the $A$ is the coefficient matrix in the
channel model \eqref{sys:h-model-general}.} Note that $\widetilde{\mathbf{z}}_{e}^{*}=0$
is an equilibrium point for $\dot{\widetilde{\mathbf{z}}}_{e}=\widetilde{\mathcal{Z}}(\widetilde{\mathbf{z}}_{e})$.
The following lemma summarizes the stability results of the virtual
system $\widetilde{\mathcal{Z}}_{e}$.
\begin{lyxLemQED}
[Stability of $\widetilde{\mathcal{Z}}_{e}$ under Quasi-Time Varying
CSI]\label{lem:Stability-z_e-under Quasi-Time Varying} Assume the
following inequality holds for all $\mathbf{h}\in\mathcal{H}$, 
\begin{equation}
\|\varphi(\mathbf{h})A\|<\kappa\min\{2M,\,-\lambda_{\max}(A)\}\label{eqn:lem-stability-time-varying_cond1_a3}
\end{equation}
where $\lambda_{\max}(A)$ denotes the largest eigenvalue of $A$,
and $M=\min\{M_{x},M_{\lambda}\}$. The joint virtual system $\dot{\widetilde{\mathbf{z}}}_{e}=\widetilde{\mathcal{Z}}(\widetilde{\mathbf{z}}_{e})$
has an exponentially stable equilibrium at $\widetilde{\mathbf{z}}_{e}^{*}=0$,
and there exists a Lyapunov function $V:\,\mathbb{R}^{n+m+q}\mapsto\mathbb{R}$
satisfying 
\begin{equation}
a_{1}\|\widetilde{\mathbf{z}}_{e}\|^{2}\leq V(\widetilde{\mathbf{z}}_{e})\leq a_{2}\|\widetilde{\mathbf{z}}_{e}\|^{2}\label{eqn:lem-stability-time-varying_V}
\end{equation}
\begin{equation}
\dot{V}(\widetilde{\mathbf{z}}_{e})=\frac{\partial V}{\partial\mathbf{z}_{e}}\widetilde{\mathcal{Z}}(\widetilde{\mathbf{z}}_{e})\leq-a_{3}\|\widetilde{\mathbf{z}}_{e}\|^{2}\label{eqn:lem-stability-time-varying_V-dot}
\end{equation}
\begin{equation}
\left\Vert \frac{\partial V}{\partial\widetilde{\mathbf{z}}_{e}}\right\Vert \leq a_{4}\|\widetilde{\mathbf{z}}_{e}\|\label{eqn:lem-stability-time-varying_pV}
\end{equation}
for some positive constants $a_{1}$, $a_{2}$, $a_{3}$ and $a_{4}>0$.\end{lyxLemQED}
\begin{proof}
Please refer to Appendix \ref{sec:App_Proof-of-Lemma_stability z_e quasi time varying}
for the proof.\end{proof}
\begin{remrk}
[Interpretation of Lemma \ref{lem:Stability-z_e-under Quasi-Time Varying}]
\label{rem:Interpretation-of-Lemma z_e quasi}From Lemma \ref{lem:Stability-z_e-under Quasi-Time Varying},
even though CSI $\mathbf{h}(t)$ is quasi-time varying, we still need
condition \eqref{eqn:lem-stability-time-varying_cond1_a3} to guarantee
globally exponential stability during the transient of $\mathbf{h}(t)$.
In fact, the parameter $\kappa$ affects the convergence rate of the
primal-dual algorithm, the eigenvalues of $A$ represent the the transient
rate of $\mathbf{h}(t)$ and the value $\|\varphi(\mathbf{h})\|$
can be considered as the sensitivity of the moving equilibrium $\widetilde{\mathbf{x}}_{e}^{*}(t)$
to the variation of $\mathbf{h}(t)$. The result of Lemma \ref{lem:Stability-z_e-under Quasi-Time Varying}
illustrates that when the transient rate of $\mathbf{h}(t)$ is slow
and the sensitivity of the moving equilibrium is small, the underlying
virtual dynamic system $\widetilde{\mathcal{Z}}_{e}$ still has global
exponential stability. This property will be used to study the stability
of the virtual system under time varying CSI.
\end{remrk}

Finally, we consider the stability result of the virtual dynamic system
$\widetilde{\mathcal{Z}}_{e}$ under the time varying CSI model in
\eqref{sys:h-model-general}. Specifically, the virtual dynamic system
is given by

\begin{eqnarray}
\widetilde{\mathcal{Z}}_{e}(u):\quad\dot{\widetilde{\mathbf{z}}}_{e} & = & \left[\begin{array}{c}
\widetilde{f}_{e}(\widetilde{\mathbf{x}}_{e};\,\mathbf{h}_{e})+\varphi(\mathbf{h}_{e}+\bar{\mathbf{h}})A\mathbf{h}_{e}\\
A\mathbf{h}_{e}
\end{array}\right]+\left[\begin{array}{c}
\varphi(\mathbf{h}_{e}+\bar{\mathbf{h}})\\
\mathbf{I}
\end{array}\right]u(t)\nonumber \\
 & = & \widetilde{\mathcal{Z}}(\widetilde{\mathbf{z}}_{e})+\Phi(\mathbf{h}_{e})u(t).\label{sys:z-u(t)}
\end{eqnarray}
Note that since $u(t)\neq0$, the CSI $\mathbf{h}(t)$ is time varying
even after the transient and its impact on the stability of the virtual
system is captured by the \emph{virtual exogenous input }$\Phi(\mathbf{h}_{e})u(t)$
in \eqref{sys:z-u(t)}. The following theorem summarizes the stability
results for $\widetilde{\mathcal{Z}}_{e}(u)$.
\begin{lyxThmQED}
[Stability of $\widetilde{\mathcal{Z}}_{e}(u)$ for Time Varying
CSI]\label{thm:Stablity of Z_e(u)} Given $\|\Phi(\mathbf{h}_{e})\|\leq\gamma$,
$\overline{\|u(t)\|^{2}}\leq\alpha^{2}$, the average trajectory $\overline{\|\widetilde{\mathbf{z}}_{e}\|^{2}}$
of the virtual dynamic system $\widetilde{\mathcal{Z}}_{e}(u)$ in
\eqref{sys:z-u(t)} satisfies 
\begin{equation}
\overline{\|\widetilde{\mathbf{z}}_{e}\|^{2}}=\lim_{T\to\infty}\frac{1}{T}\int_{0}^{T}\|\widetilde{\mathbf{z}}_{e}(t)\|^{2}dt\le\frac{a_{4}^{2}\gamma^{2}}{a_{3}^{2}}\alpha^{2}\label{eqn:stability region of z_e(t)}
\end{equation}
where $\overline{\|u(t)\|^{2}}=\lim_{T\to\infty}\frac{1}{T}\int_{0}^{T}\|u(t)\|^{2}dt$.\end{lyxThmQED}
\begin{proof}
Please refer to Appendix \ref{sec:App_Proof-of-Theorem_stability_z_e}
for the proof.
\end{proof}

Note that $\overline{\|u(t)\|^{2}}$ denotes the \emph{average exogenous
excitation} of $u(t)$ to the virtual system $\widetilde{\mathcal{Z}}_{e}$
and $\overline{\|\widetilde{\mathbf{z}}_{e}\|^{2}}$ denotes the \emph{average
tracking error }of the system. Based on the result in Theorem \ref{thm:Stablity of Z_e(u)},
the following corollary summarizes the convergence performance of
the primal-dual algorithm for a strongly concave-convex saddle point
problem in time varying CSI.
\begin{lyxCorQED}
\label{cor:Convergence-strongly}\emph{(Convergence Performance of
Primal-Dual Algorithms for Strongly Concave-Convex Saddle Point Problems):}
Suppose $\|\varphi(\mathbf{h})\|\leq\gamma_{0}$ for all $\mathbf{h}\in\mathcal{H}$
and $\overline{\|u(t)\|^{2}}\leq\alpha^{2}$, the average tracking
error $\overline{\|\widetilde{\mathbf{x}}_{e}\|^{2}}$ of the primal-dual
algorithm under time varying CSI satisfies $\overline{\|\widetilde{\mathbf{x}}_{e}\|^{2}}=\lim_{T\to\infty}\frac{1}{T}\int_{0}^{T}\|\widetilde{\mathbf{x}}_{e}(t)\|^{2}dt\leq\frac{a_{4}^{2}(\gamma_{0}^{2}+1)}{a_{3}^{2}}\alpha^{2}$.\end{lyxCorQED}
\begin{proof}
Since $\|\Phi(\mathbf{h}_{e})\|=\sqrt{\|\varphi(\mathbf{h}_{e}+\bar{\mathbf{h}})\|^{2}+1}\leq\sqrt{\gamma_{0}^{2}+1}$,
by Theorem \ref{thm:Stablity of Z_e(u)}, the average trajectory of
the corresponding virtual dynamic system $\widetilde{\mathcal{Z}}_{e}(u)$
for the primal-dual iterative algorithm converges to $\overline{\|\widetilde{\mathbf{z}}_{e}\|^{2}}=\lim_{T\to\infty}\frac{1}{T}\int_{0}^{T}\|\widetilde{\mathbf{z}}_{e}(t)\|^{2}dt\le\frac{a_{4}^{2}(\gamma_{0}^{2}+1)}{a_{3}^{2}}\alpha^{2}$.
As $\|\widetilde{\mathbf{x}}_{e}\|\leq\|\widetilde{\mathbf{z}}_{e}\|$
the average trajectory $\overline{\|\widetilde{\mathbf{x}}_{e}(t)\|^{2}}$
of the error process $\widetilde{\mathcal{X}}_{e}$ converges to $\frac{a_{4}^{2}(\gamma_{0}^{2}+1)}{a_{3}^{2}}\alpha^{2}$,
and equivalently, the average tracking error $\overline{\|\widetilde{\mathbf{x}}_{e}\|^{2}}$
of the primal-dual algorithm satisfies $\lim_{T\to\infty}\frac{1}{T}\int_{0}^{T}\|\widetilde{\mathbf{x}}_{e}(t)\|^{2}dt\leq\frac{a_{4}^{2}(\gamma_{0}^{2}+1)}{a_{3}^{2}}\alpha^{2}$.
This completes the proof of Corollary \ref{cor:Convergence-strongly}. 
\end{proof}

Note that the average tracking error $\overline{\|\widetilde{\mathbf{x}}_{e}\|^{2}}$
depends on $\alpha^{2}$, $\gamma_{0}$ and $a_{4}/a_{3}$ which represents
how fast the CSI changes, how sensitive the equilibrium $\widetilde{\mathbf{x}}^{*}(t)$
is w.r.t. the change of CSI $\mathbf{h}(t)$, and how fast the algorithm
converges, respectively. For instance, the average tracking error
$\overline{\|\widetilde{\mathbf{x}}_{e}\|^{2}}$ will be smaller if
the CSI is slowly changing (i.e. $\overline{\|u(t)\|^{2}}$ is small)
or the convergence rate of the underlying primal-dual algorithm (under
time invariant CSI) is fast (i.e. smaller $a_{4}/a_{3}$).

\subsection{Numerical Example\label{sub:Numerical-Example_JammingGame}}

We shall illustrate the application of Theorem \ref{thm:Stablity of Z_e(u)}
using a numerical example from Section \ref{subsub:App_Transmission_Jamming}.
Consider a 2-antenna system; the fading channels are given by $\mathbf{H}_{1},\mathbf{H}_{2}\in\mathbb{C}^{2\times2}$,
and $\mathbf{h}=\mbox{vec}\left([\mathbf{H}_{1}\,\mathbf{H}_{2}]\right)\in\mathbb{C}^{8}$.
We choose the channel model in \eqref{sys:h-model-general} as $A=-aI$,
$\bar{\mathbf{h}}=\mathbf{1}$ and $u(t)=\sqrt{2a}w(t)$, where $w(t)$
is a zero-mean unit-variance white Gaussian process. This corresponds
to a standard AR Gaussian fading process. The parameter $a$ controls
the time-correlation of the CSI and the CSI $\mathbf{h}(t)$ has unity
variance. The primal-dual algorithm for the specific example is given
by 
\begin{equation}
\left[\begin{array}{c}
\frac{d\mbox{vec}(\mathbf{Q})}{dt}\\
\frac{d\mbox{vec}(\mathbf{Z})}{dt}
\end{array}\right]=\left[\begin{array}{c}
\left\{ \left(\frac{\partial}{\partial\mbox{vec}^{T}(\mathbf{Q})}C\left(\mathbf{Q},\,\mathbf{Z};\,\mathbf{h}(t)\right)\right)^{T}\right\} _{\mathcal{Q}}\\
\left\{ \left(-\frac{\partial}{\partial\mbox{vec}^{T}(\mathbf{Z})}C\left(\mathbf{Q},\,\mathbf{Z};\,\mathbf{h}(t)\right)\right)^{T}\right\} _{\mathcal{J}}
\end{array}\right]\triangleq\widetilde{f}(\widetilde{\mathbf{x}};\,\mathbf{h}(t))\label{eqn:exp2-p-d-alg}
\end{equation}
where $\mathcal{Q}=\{\mathbf{Q}\in\mathbb{C}^{2\times2}:\mathbf{Q}\succeq0,\,\mbox{tr}(\mathbf{Q})\leq P_{T}\}$
and $\mathcal{J}=\{\mathbf{Z}\in\mathbb{C}^{2\times2}:\mathbf{Z}\succeq0,\,\mbox{tr}(\mathbf{Z})\leq P_{J}\}$
are the feasible sets for the transmission covariance matrices, projections
$\{\centerdot\}_{\mathcal{Q}}$ and $\{\centerdot\}_{\mathcal{J}}$
are to restrict the searching directions within the feasible sets
$\mathcal{Q}$ and $\mathcal{J}$, respectively, and $\widetilde{\mathbf{x}}=[\mbox{vec}^{T}(\mathbf{Q})\;\mbox{vec}^{T}(\mathbf{Z})]^{T}$.
In this numerical example, the power constraints are set to be $P_{T}=P_{J}=10$.

Correspondingly, the error dynamic system $\widetilde{\mathcal{X}}_{e}$
and the virtual dynamic system $\widetilde{\mathcal{Z}}_{e}(u)$ is
given by \eqref{sys:error-dynamic-system} and \eqref{sys:z-u(t)}
respectively, with $\widetilde{f}(\widetilde{\mathbf{x}};\,\mathbf{h}(t))$
defined in \eqref{eqn:exp2-p-d-alg}. The equilibrium point is defined
as $\widetilde{f}(\widetilde{\mathbf{x}}^{*};\,\mathbf{h}(t))=\mathbf{0}$
and the function $\varphi(\mathbf{h})\triangleq\frac{\partial}{\partial\mathbf{h}}\widetilde{\mathbf{x}}^{*}(\mathbf{h})=\left(\frac{\partial}{\partial\mathbf{\widetilde{\mathbf{x}}}}\widetilde{f}(\widetilde{\mathbf{x}}^{*};\,\mathbf{h})\right)^{-1}\frac{\partial}{\partial\mathbf{h}}\widetilde{f}(\widetilde{\mathbf{x}}^{*};\,\mathbf{h})$.

It can be numerically verified that $\|\varphi(\mathbf{h})\|\leq11.22$
for all possible $\mathbf{h}\in\mathcal{H}\subseteq\mathbb{C}^{8}$,
and the condition \eqref{eqn:lem-stability-time-varying_cond1_a3}
in Lemma \ref{lem:Stability-z_e-under Quasi-Time Varying} is satisfied
by choosing $\kappa=1\mbox{ sec}^{-1}$. Hence by Theorem \ref{thm:Stablity of Z_e(u)},
we conclude that the average tracking error $\frac{1}{T}\int_{0}^{T}\|\widetilde{\mathbf{x}}(t)-\widetilde{\mathbf{x}}^{*}(t)\|^{2}dt$
converges to $a_{4}^{2}\gamma^{2}\alpha^{2}/a_{3}^{2}$ with $\gamma=\sqrt{\|\varphi(\mathbf{h})\|^{2}+1}\leq11.26$,
$\alpha^{2}=16a$ and $a_{4}/a_{3}=0.222$ (when the CSI parameter
$a=0.02\mbox{ ms}^{-1}$).

\section{Convergence Analysis of Degraded Saddle Point Problems under Time
Varying CSI\label{sec:Convergence-Analsyiss-Degraded}}

In Section \ref{sec:Convegenece-Analysis-unconstrained}, we have
focused on the case where the objective function $\mathcal{L}(\mathbf{x},\,\lambda)$
is strongly concave in $\mathbf{x}$ and strongly convex in $\lambda$.
This property provides exponential convergence rate to the saddle
point $(\mathbf{x}^{*},\,\mathbf{\lambda}^{*})$ of \eqref{eqn:min-max}.
In this section we shall extend the results to deal with the case
when one of the Hessians is only semidefinite. Without loss of generality,
we assume $\nabla_{x}^{2}\mathcal{L}\preceq-M_{x}\mathbf{I}$ for
$M_{x}>0$ and $\nabla_{\lambda}^{2}\mathcal{L}\succeq0$, and this
corresponds to the case where the objective function $\mathcal{L}(\mathbf{x},\,\mathbf{\lambda})$
is strongly concave in $\mathbf{x}$ and only convex in $\mathbf{\lambda}$.
We call such a saddle point problem the \emph{degraded saddle point
problem} and the associated virtual dynamic systems the \emph{degraded
virtual dynamic systems}. \textcolor{black}{Examples of degraded
saddle point problems include the primal-dual iterations (single time
scale) of convex optimization problems. For instance, the Lagrangian
function \eqref{eqn:pro1-Lagrangian} of the NUM problem \eqref{eqn:pro1-formulation}
is strongly concave in the primal variables $\mathbf{r}$ and $\mathbf{p}$
but only convex in the LM $\lambda$. }

\subsection{Stability Analysis of the Degraded Virtual Dynamic System}

Similarly, we first consider the case when the CSI is time invariant
($\dot{\mathbf{h}}=0$). For degraded saddle point problems, the exponential
stability property of the associated virtual dynamic system $\widetilde{\mathcal{X}}_{e}$
in Lemma \ref{lem:stability of x_e with time-invariant CSI} does
not hold. We have a \emph{partial exponential stability }property
summarized in the following.
\begin{lyxLemQED}
[Partial Exponential Stability of $\widetilde{\mathcal{X}}_{e}$
with Time Invariant CSI]\label{lem:Partial Exponential Stability of x_e}
When the CSI is time invariant, the equilibrium point $\widetilde{\mathbf{x}}_{e}^{*}=(\mathbf{x}_{e}^{*},\,\mathbf{\lambda}_{e}^{*})=0$
of the virtual dynamic system $\widetilde{\mathcal{X}}_{e}$ in \eqref{sys:error-dynamic-system}
is \emph{partially exponentially stable} w.r.t. $\mathbf{x}_{e}^{*}$
for all $\mathbf{h}\in\mathcal{H}$, i.e., there exists a Lyapunov
function $V_{x}:\,\mathbb{R}^{n+m}\times\mathbb{R}^{q}\mapsto\mathbb{R}$
defined on the virtual state $\widetilde{\mathbf{x}}_{e}$ that satisfies
\begin{equation}
c_{1}\|\widetilde{\mathbf{x}}_{e}\|^{2}\leq V_{x}(\widetilde{\mathbf{x}}_{e})\leq c_{2}\|\widetilde{\mathbf{x}}_{e}\|^{2}\label{eqn:lem-partial-exp_y_V}
\end{equation}
\begin{equation}
\dot{V}_{x}(\widetilde{\mathbf{x}}_{e})=\frac{\partial V}{\partial\mathbf{\widetilde{\mathbf{x}}_{e}}}\widetilde{f}_{e}(\widetilde{\mathbf{x}}_{e};\,\mathbf{h})\leq-c_{3}\|\mathbf{x}_{e}\|^{2}\label{eqn:lem-partial-exp_y-V-dot}
\end{equation}
\begin{equation}
\left\Vert \frac{\partial V_{x}}{\partial\widetilde{\mathbf{x}}_{e}}\right\Vert \leq c_{4}\|\widetilde{\mathbf{x}}_{e}\|.\label{eqn:lem-partial-exp_y_pV}
\end{equation}

\end{lyxLemQED}

The proof is similar to that in Appendix \ref{sec:App_Proof-of-Lemma_stability_x_e},
by simply taking a Lyapunov function $V_{x}(\widetilde{\mathbf{x}}_{e})=\frac{1}{2\kappa}\widetilde{\mathbf{x}}_{e}^{T}\widetilde{\mathbf{x}}_{e}$,
and letting $M_{\lambda}=0$. For this Lyapunov function, $c_{1}=c_{2}=\frac{1}{2\kappa}$,
$c_{3}=2M_{x}$ and $c_{4}=1/\kappa$. Note that, instead of possessing
exponential stability on the joint state $\widetilde{\mathbf{x}}_{e}^{*}=(\mathbf{x}_{e}^{*},\,\lambda_{e}^{*})$,
the degraded virtual dynamic system just has exponential stability
on $\mathbf{x}_{e}^{*}$ as shown in \eqref{eqn:lem-partial-exp_y-V-dot}. 

Similarly, we study the stability of the virtual dynamic system $\dot{\widetilde{\mathbf{x}}}_{e}=\widetilde{f}_{e}(\widetilde{\mathbf{x}}_{e};\,\mathbf{h}(t))-\varphi(\mathbf{h})\dot{\mathbf{h}}(t)$
with a quasi-time varying CSI $\mathbf{h}(t)$ for $u(t)=0$. Since
we have exponential stability only on $\mathbf{x}_{e}^{*}$ but not
on $\mathbf{\lambda}_{e}^{*}$, we define a partial state as $\mathbf{z}_{e}=(\mathbf{x}_{e},\,\mathbf{h}_{e})$.
As a result, the trajectory of the partial state $\mathbf{z}_{e}$
satisfies the following virtual dynamic system $\mathcal{Z}_{e}$
: 
\begin{equation}
\mathcal{Z}_{e}:\quad\dot{\mathbf{z}}_{e}=\mathcal{Z}(\mathbf{z}_{e}):=\left[\begin{array}{c}
f_{e}(\mathbf{x}_{e},\,\lambda_{e}=0;\,\mathbf{h}_{e}+\bar{\mathbf{h}})+\varphi_{x}(\mathbf{h}_{e}+\bar{\mathbf{h}})A\mathbf{h}_{e}\\
A\mathbf{h}_{e}
\end{array}\right]\label{sys:degraded z(x_e,h)}
\end{equation}
where $f_{e}(\mathbf{x}_{e},\,\lambda_{e};\,\mathbf{h}(t))=\left(\frac{\partial}{\partial\mathbf{x_{e}}}\mathcal{L}(\mathbf{x}_{e}+\mathbf{x}^{*},\,\lambda_{e}+\lambda^{*};\,\mathbf{h}(t))\right)^{T}$
and $\varphi_{x}(\mathbf{h})=\frac{\partial}{\partial\mathbf{h}}\mathbf{x}^{*}(\mathbf{h})=\frac{\partial}{\partial\mathbf{h}}\psi_{x}(\mathbf{h})$.
The following lemma summarizes the stability results for the virtual
system $\dot{\mathbf{z}}_{e}=\mathcal{Z}(\mathbf{z}_{e})$.
\begin{lyxLemQED}
[Partial Exponential Stability of $\mathcal{Z}_{e}$ under Quasi-Time
Varying CSI]\label{lem:Partial-Exponential-Stability of z_e quasi-time}
Suppose the following inequality holds for all $\mathbf{h}\in\mathcal{H}$:
\begin{equation}
\|\varphi_{x}(\mathbf{h})A\|<\kappa\min\{2M_{x},\,-\lambda_{\max}(A)\}\label{eqn:lem-partial-exp_cond1}
\end{equation}
where $\lambda_{\max}(A)$ denotes the largest eigenvalue of $A$.
Then the system $\mathcal{Z}_{e}$ in \eqref{sys:degraded z(x_e,h)}
is \emph{partially exponentially stable}, i.e., there exists a Lyapunov
function for the joint state $\mathbf{z}_{e}$ satisfying 
\begin{equation}
a_{1}\|\mathbf{z}_{e}\|^{2}\leq V(\mathbf{\mathbf{z}_{e}})\leq a_{2}\|\mathbf{z}_{e}\|^{2}\label{eqn:lem-partial-exp_z_V}
\end{equation}
\begin{equation}
\dot{V}(\mathbf{z}_{e})=\frac{\partial V}{\partial\mathbf{z}_{e}}Z(\mathbf{z}_{e})\leq-a_{3}\|\mathbf{z}_{e}\|^{2}\label{eqn:lem-partial-exp_z_V-dot}
\end{equation}
\begin{equation}
\left\Vert \frac{\partial V}{\partial\mathbf{z}_{e}}\right\Vert \leq a_{4}\|\mathbf{z}_{e}\|.\label{eqn:lem-partial-exp_z_pV}
\end{equation}
\end{lyxLemQED}
\begin{proof}
Please refer to Appendix \ref{sec:App_Proof-of-Lemma_partial stability z_e quasi}
for the proof.
\end{proof}

Lemma \ref{lem:Partial-Exponential-Stability of z_e quasi-time} suggests
that as long as the transient of quasi-time varying $\mathbf{h}(t)$
is not changing too fast (i.e. $A$ has small eigenvalues) and the
sensitivity of the primal part of the equilibrium $\mathbf{x}_{e}^{*}$
w.r.t. the change of $\mathbf{h}(t)$ is small (i.e. small $\|\varphi_{x}(\mathbf{h})\|$),
the virtual dynamic system $\mathcal{Z}_{e}$ still possesses globally
exponential stability on the partial state $\mathbf{z}_{e}=(\mathbf{x}_{e},\,\mathbf{h}_{e})$.

Finally, we consider the stability result of the degraded virtual
dynamic system under the time varying CSI model in \eqref{sys:h-model-general}.
The degraded virtual dynamic system $\mathcal{Z}_{e}(u)$ is given
by $ $
\begin{eqnarray}
\mathcal{Z}_{e}(u):\quad\dot{\mathbf{z}}_{e} & = & \left[\begin{array}{c}
f_{e}(\mathbf{x}_{e},\,\lambda_{e}=0;\,\mathbf{h})+\varphi_{x}(\mathbf{h}_{e}+\bar{\mathbf{h}})A\mathbf{h}_{e}\\
A\mathbf{h}_{e}
\end{array}\right]+\left[\begin{array}{c}
\varphi_{x}(\mathbf{h}_{e}+\bar{\mathbf{h}})\\
\mathbf{I}
\end{array}\right]u(t)\nonumber \\
 & = & \mathcal{Z}(\mathbf{z}_{e})+\Phi_{x}(\mathbf{h}_{e})u(t).\label{sys:parital joint z-u(t)}
\end{eqnarray}
The stability result is summarized in the following theorem.
\begin{lyxThmQED}
[Stability of $\mathcal{Z}_{e}(u)$ for Time Varying CSI]\label{thm:Stability-of parital joint z_e(u) time varying}
Given $\|\Phi_{x}(\mathbf{h}_{e})\|\leq\gamma_{x}$, and $\overline{\|u(t)\|^{2}}\leq\alpha^{2}$,
the average trajectory $\overline{\|\mathbf{z}_{e}\|^{2}}$ of the
degraded virtual dynamic system $\mathcal{Z}_{e}(u)$ satisfies 
\[
\overline{\|\mathbf{z}_{e}\|^{2}}=\frac{1}{T}\int_{0}^{T}\|\mathbf{z}_{e}(t)\|^{2}dt\leq\frac{a_{4}^{2}\gamma_{x}^{2}}{a_{3}^{2}}\alpha^{2}
\]
where $\overline{\|u(t)\|^{2}}=\lim_{T\to\infty}\frac{1}{T}\int_{0}^{T}\|u(t)\|^{2}dt$.
\end{lyxThmQED}

The proof can be derived similarly from that of Theorem \ref{thm:Stablity of Z_e(u)}
using the property of Lemma \ref{lem:Partial-Exponential-Stability of z_e quasi-time}.
As a result, we summarize the convergence performance of the primal-dual
iterative algorithm for a degraded saddle point problem in the following
corollary.
\begin{lyxCorQED}
\label{cor:convergence-perf-degraded saddle point}\emph{(Convergence
Performance of Primal-Dual Algorithms for Degraded Saddle Point Problems):
}Suppose $\varphi_{x}(\mathbf{h})\leq\bar{\gamma}_{x}$ for all $\mathbf{h}\in\Gamma$
and $\overline{\|u(t)\|^{2}}\leq\alpha^{2}$, the average tracking
error $\overline{\|\mathbf{x}_{e}\|^{2}}$ for the primal-dual algorithm
under time varying CSI satisfies $\overline{\|\mathbf{x}_{e}\|^{2}}=\lim_{T\to\infty}\frac{1}{T}\int_{0}^{T}\|\mathbf{x}_{e}(t)\|^{2}dt\leq\frac{a_{4}^{2}(\bar{\gamma}_{x}^{2}+1)}{a_{3}^{2}}\alpha^{2}$. 
\end{lyxCorQED}

The proof can be derived similarly from that of Corollary \ref{cor:Convergence-strongly},
by applying Theorem \ref{thm:Stability-of parital joint z_e(u) time varying}
and using $\|\Phi_{x}(\mathbf{h}_{e})\|=\sqrt{\|\varphi_{x}(\mathbf{h}_{e}+\bar{\mathbf{h}})\|^{2}+1}\leq\sqrt{\bar{\gamma}_{x}^{2}+1}$
and $\|\mathbf{x}_{e}\|\leq\|\mathbf{z}_{e}\|$. 

Together with Lemma \ref{lem:Partial-Exponential-Stability of z_e quasi-time},
the above theorem and corollary establish sufficient conditions on
the convergence performance of the a primal-dual algorithm for a degraded
saddle point problem under time varying CSI. It shows that once the
corresponding virtual dynamic system $\dot{\mathbf{z}}_{e}=\mathcal{Z}_{u}(\mathbf{z}_{e},\, u(t))$
is exponentially stable at $\mathbf{z}_{e}^{*}=0$ for $u(t)=0$,
the system is stable for general $u(t)$ as long as $\overline{\|u(t)\|^{2}}$
is bounded. Correspondingly, the tracking error of the primal-dual
algorithm under time varying CSI is bounded and scaled according to
$\mathcal{O}\left(\overline{\|u(t)\|^{2}}\right)$. 

Note that, due to the degraded saddle point problem in the dual variable
$\mathbf{\lambda}$, the average error bound $\frac{a_{4}^{2}(\bar{\gamma}_{x}^{2}+1)}{a_{3}^{2}}\alpha^{2}$
in Corollary \ref{cor:convergence-perf-degraded saddle point} does
not include the dual variable $\mathbf{\lambda}_{e}$, and hence we
cannot bound the tracking error of the dual variables. This is because
due to the semidefinite Hessian matrix $\nabla_{\lambda}\mathcal{L}\succeq0$,
the dual variable $\lambda$ in the primal-dual iterative algorithm
\eqref{eqn:alg_Arrow-gradient-h(t)-1}-\eqref{eqn:alg_Arrow-gradient-h(t)-2}
may not converge exponentially fast to the saddle point $\lambda^{*}$.
However, this result is still meaningful because, in many applications
such as constrained optimizations, we are mainly concerned about the
behavior of the primal variable $\mathbf{x}$. We shall illustrate
with an example in the next section.

\subsection{Numerical Example\label{sub:Numerical example - NUM}}

We illustrate an application of Theorem \ref{thm:Stability-of parital joint z_e(u) time varying}
using a numerical example in Section \ref{subsub:App_NUM}. Consider
a simple wireless network with 3 nodes depicted in Fig. \ref{fig:Multihop wireless network},
where the collection of traffic flows is given by $\mathcal{C}=\{(1,2),\,(1,3),\,(2,3)\}$,
and the sets of links are $L(1,2)=\{1\}$, $L(1,3)=\{1,2\}$ and $L(2,3)=\{2\}$.
$P_{t1}$ and $P_{t2}$ denote the total power allocated to link 1
and link 2, respectively, and $h_{1}$ and $h_{2}$ are the corresponding
channel gains. Let $\mathbf{h}=[h_{1}\; h_{2}]^{T}$. Our CSI dynamic
is modeled in \eqref{sys:h-model-general}, with $A=-aI$ and $u(t)=\sqrt{2a}w(t)$
where $w(t)$ is a zero-mean unit-variance white Gaussian process.
The variance of the CSI $\mathbf{h}(t)$ is normalized to unity and
the parameter $a$ controls the time-correlation of the CSI. The average
receiving SNR is normalized to 10 dB. We consider $U_{sd}(r_{sd})=\log(1+r_{sd})$
as the cost function, and the link capacity function is given by $c_{l}(P_{tl},\, h_{l})=\log(1+h_{l}^{2}P_{tl}),\,\forall l=1,2$.
The Lagrangian saddle point problem for this specific example is given
by $ $
\[
\min_{\lambda\succeq0}\max_{\mathbf{r}}\mathcal{L}(\mathbf{r},\,\mathbf{\lambda})=\sum_{(s,d)\in\mathcal{C}}\log(1+r_{sd})-\sum_{l=1,2}\lambda_{l}\left(\sum_{(s,d):l\in L(s,d)}r_{sd}-\log(1+h_{l}^{2}P_{tl})\right)+\sum_{(s,d)\in\mathcal{C}}\mathbf{\lambda}_{r}^{(sd)}r_{sd}.
\]

The error dynamic system and the degraded virtual dynamic system are
given by \eqref{sys:error-dynamic-system} and \eqref{sys:degraded z(x_e,h)},
respectively. Specifically, let $\mathbf{r}=[r_{12}\; r_{13}\; r_{23}]^{T}$
be the primal variable and $\bm{\lambda}$ be the collection of the
dual variables. Then $\widetilde{\mathbf{x}}_{e}=\widetilde{\mathbf{x}}-\widetilde{\mathbf{x}}^{*}$,
$\widetilde{\mathbf{x}}=(\mathbf{r},\,\mathbf{\lambda})$, $\mathbf{z}_{e}=(\mathbf{r},\,\mathbf{h})$,
$\Phi_{\mathbf{r}}(\mathbf{h})=[\varphi_{\mathbf{r}}(\mathbf{h});\, I]$,
and $\varphi(\mathbf{h})=\frac{\partial}{\partial\mathbf{h}}\mathbf{\widetilde{x}}^{*}(\mathbf{h})=\left(\frac{\partial F}{\partial\widetilde{\mathbf{x}}}\right)^{-1}\frac{\partial F}{\partial\mathbf{h}}$
which is derived by the implicit function theorem, where the implicit
function $F(\mathbf{r}^{*},\,\lambda^{*})=0$ is given by the KKT
conditions 
\[
F(\mathbf{r}^{*},\,\lambda^{*})=\left[\begin{array}{c}
\nabla_{\mathbf{r}}\mathcal{L}(\mathbf{r}^{*},\,\lambda^{*})\\
\lambda_{l}^{*}\left(\sum r_{sd}^{*}-c_{l}(P_{tl},\, h_{l})\right)\\
\lambda_{r}^{(sd)*}r_{sd}^{*}
\end{array}\right]=\mathbf{0}.
\]

Numerical results suggest that $\|\varphi_{\mathbf{r}}(\mathbf{h})\|\leq2.2$,
and hence the constants in Lemma \ref{lem:Partial Exponential Stability of x_e}
can be chosen as $\kappa=0.5\mbox{ sec}^{-1}$, such that condition
\eqref{eqn:lem-partial-exp_cond1} in Lemma \ref{lem:Partial-Exponential-Stability of z_e quasi-time}
is satisfied. Thus we can apply Corollary \ref{cor:convergence-perf-degraded saddle point},
which shows that the average tracking error for the rate allocation
variables $\overline{\|\mathbf{r}_{e}\|^{2}}=\lim_{T\to\infty}\frac{1}{T}\int_{0}^{T}\|\mathbf{r}_{e}(t)\|^{2}dt\leq\frac{a_{4}^{2}}{a_{3}^{2}}\gamma_{r}^{2}\alpha^{2}$,
where $\gamma_{r}\leq\sqrt{\|\varphi_{r}(\mathbf{h})\|^{2}+1}\leq2.42$,
$\alpha^{2}=4a$, and $a_{4}/a_{3}=1.69$ (when the CSI parameter
$a=0.04\mbox{ ms}^{-1}$).

\section{Adaptive Compensation for the Primal Dual Algorithms in Time Varying
Channels\label{sec:Adaptive-Proposed-alg}}

In the previous sections, we have analyzed the convergence behavior
of the primal-dual algorithms under time varying CSI. In this section,
we consider modifying the primal-dual iterations to reduce the \emph{tracking
error }in time varying CSI. To this end, we shall introduce a \emph{compensation
term }in the iteration to offset the \emph{exogenous excitation }to
the virtual dynamic system in \eqref{sys:z-u(t)} and \eqref{sys:parital joint z-u(t)}.
In the following, we shall first discuss the construction of the compensation
term. Based on that, we shall obtain a low complexity distributive
implementation for the compensation method.

\subsection{Adaptive Compensation for the Primal Dual Algorithm}

\begin{figure}
\begin{centering}
\includegraphics[scale=0.45]{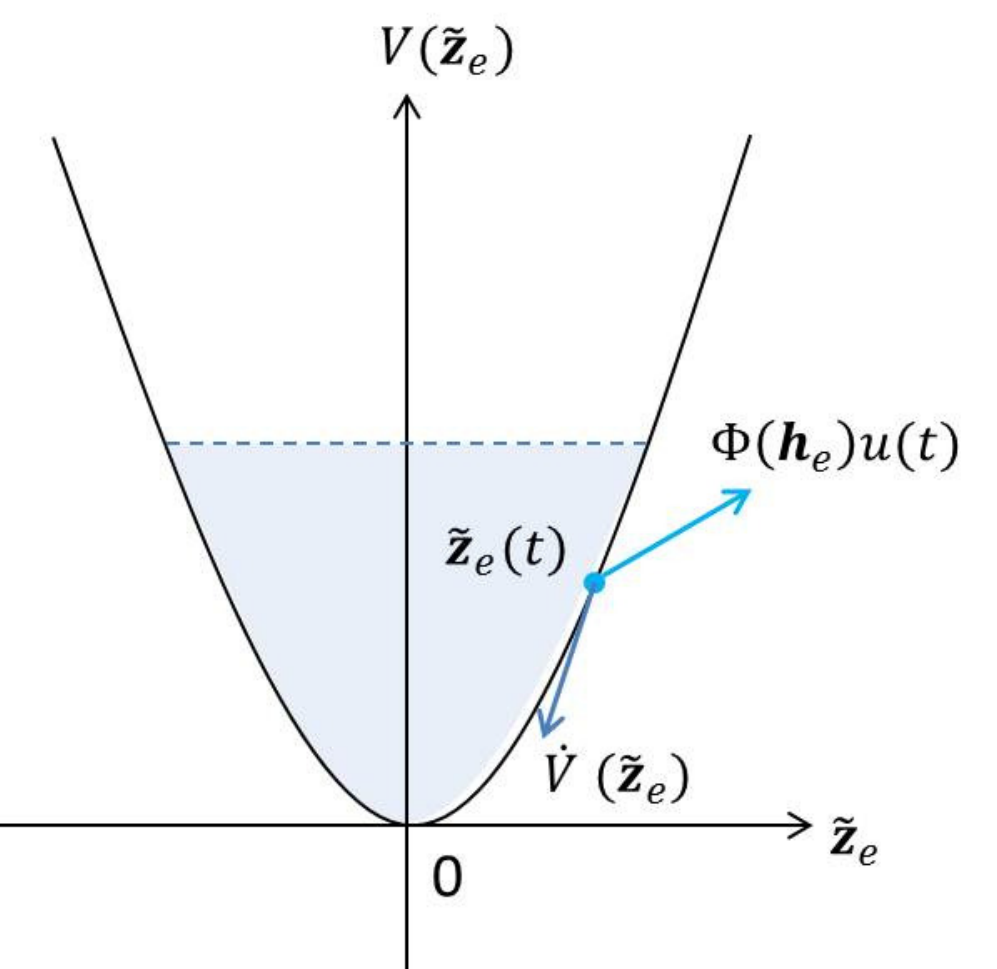}
\par\end{centering}

\caption{\label{fig:Engergy-Disturbance}An illustration of the excitation
due to the exogenous input $\Phi(\mathbf{h}_{e})u(t)$ for the equivalent
virtual dynamic system $\widetilde{\mathcal{Z}}_{e}$. We can visualize
the Lyapunov function $V(\widetilde{\mathbf{z}}_{e})$ as an energy
function. Without the time-varying channel effect ($u(t)=0$), the
algorithm trajectory will eventually converges to the equilibrium
$\widetilde{\mathbf{z}}_{e}^{*}=0$ (corresponding to exponential
convergence of the primal-dual iterations in \eqref{eqn:alg_Arrow-gradient-h(t)-1}-\eqref{eqn:alg_Arrow-gradient-h(t)-2})
because there is no exogenous excitation energy applied to the system.
With the time-varying channels, there is an uncertainty region (shaded
region) where the system state will converge to (corresponds to the
convergence region $\widetilde{D}_{e}$ in Corollary \ref{cor:Convergence-strongly}
and $D_{e}$ in Corollary \ref{cor:convergence-perf-degraded saddle point}).
The region is bounded and the size depends on the \textquotedbl{}energy\textquotedbl{}
applied to the system due to the exogenous input $\Phi(\mathbf{h}_{e})u(t)$. }

\end{figure}

We have shown in Theorem \ref{thm:Stablity of Z_e(u)} and Theorem
\ref{thm:Stability-of parital joint z_e(u) time varying} that the
average tracking error of a primal-dual algorithm under time varying
CSI is $\mathcal{O}(\alpha^{2}\gamma^{2})$ where $\alpha^{2}\gamma^{2}$
is the square average of the norm of the exogenous input $\Phi(\mathbf{h}_{e})u(t)$
to the virtual dynamic system $\widetilde{\mathcal{Z}}_{e}$. Consider
a particle trapped in an \emph{energy well }given by the Lyapunov
function%
\footnote{The Lyapunov function can be interpreted as the \emph{energy function
}of the system state $\widetilde{\mathbf{z}}_{e}$.%
} $V(\widetilde{\mathbf{z}}_{e})$. The presence of the exogenous input
term $\Phi(\mathbf{h}_{e})u(t)$ to the system $\widetilde{\mathcal{Z}}_{e}$
excites the particle and moves it away from the equilibrium position,
and how far the particle can be moved depends on the \emph{exogenous
excitation energy }induced by $u(t)$. This phenomenon is illustrated
in Fig. \ref{fig:Engergy-Disturbance}. As a result, one way to reduce
the tracking error is to compensate the disturbance from the exogenous
input $\Phi(\mathbf{h}_{e})u(t)$. We thus introduce a compensation
term $\widehat{\Phi}(\widetilde{\mathbf{z}}_{e})u(t)$ to the virtual
dynamic system $\widetilde{\mathcal{Z}}_{e}(u)$ in \eqref{sys:z-u(t)}
as follows, 
\begin{equation}
\widetilde{\mathcal{Z}}_{e}(\widehat{u}):\quad\dot{\widetilde{\mathbf{z}}}_{e}=\widetilde{\mathcal{Z}}(\widetilde{\mathbf{z}}_{e})+\Phi(\mathbf{h}_{e})u(t)-\widehat{\Phi}(\mathbf{\widetilde{z}}_{e})u(t)\label{sys:alg_z_u(t)}
\end{equation}
where $\widehat{\Phi}(\mathbf{\widetilde{z}}_{e})=\left[\begin{array}{c}
\widehat{\varphi}(\mathbf{\widetilde{x}}_{e},\,\mathbf{h}_{e})\\
\mathbf{I}
\end{array}\right]$ is the compensation term to be derived. Obviously, if we could set
$\widehat{\Phi}(\widetilde{\mathbf{z}}_{e})=\Phi(\mathbf{h}_{e})$,
the impact of the exogenous input $\Phi(\mathbf{h}_{e})u(t)$ can
be totally suppressed and the virtual dynamic system $\widetilde{\mathcal{Z}}_{e}(\widehat{u})$
will be free from exogenous excitation. However, since we do not have
closed form expression for the saddle point $\widetilde{\mathbf{x}}^{*}(\mathbf{h})$,
$\Phi(\mathbf{h}_{e})$ cannot be obtained during the iteration. 

On the other hand, suppose that there exists a function $F:\mathbb{R}^{n}\times\mathbb{R}_{+}^{m}\mapsto\mathbb{R}^{n+m}\in\mathcal{C}^{1}$
such that $F(\widetilde{\mathbf{x}}^{*};\,\mathbf{h})=0$ and $\frac{\partial}{\partial\mathbf{\widetilde{x}}}F(\widetilde{\mathbf{x}}^{*};\,\mathbf{h})$
is non-singular for all $\widetilde{\mathbf{x}}^{*}(\mathbf{h})$
and $\mathbf{h}\in\mathcal{H}\subset\mathbb{R}^{q}$. Using the implicit
function theorem, the function $\Phi(\mathbf{h}_{e})$ in \eqref{sys:alg_z_u(t)}
is given by 
\begin{equation}
\Phi(\mathbf{h}_{e})=\left[\begin{array}{c}
-\left[\frac{\partial}{\partial\mathbf{\widetilde{x}}}F(\widetilde{\mathbf{x}}^{*}(\mathbf{h}_{e}+\bar{\mathbf{h}});\,\mathbf{h}_{e}+\bar{\mathbf{h}})\right]^{-1}\frac{\partial}{\partial\mathbf{h}}F(\widetilde{\mathbf{x}}^{*}(\mathbf{h}_{e}+\bar{\mathbf{h}});\,\mathbf{h}_{e}+\bar{\mathbf{h}})\\
\mathbf{I}
\end{array}\right].\label{eqn:Phi(he)}
\end{equation}

Note that the function $F(\centerdot)$ can be found by the optimality
conditions of the saddle point problem. For example, it can be chosen
as $\widetilde{f}(\mathbf{\widetilde{x}}^{*};\,\mathbf{h}(t))=0$
by the definition of equilibrium point (Definition \ref{def:Equilibrium-Point}),
where $\widetilde{f}(\centerdot)$ is given in \eqref{eqn:vector field for a virtual dynamic sys}.
Using \eqref{eqn:Phi(he)}, we can estimate $\Phi(\mathbf{h}_{e})$
by 

\begin{equation}
\widehat{\Phi}(\widetilde{\mathbf{z}}_{e})=\left[\begin{array}{c}
-\left(\frac{\partial}{\partial\mathbf{\widetilde{x}}}F(\widetilde{\mathbf{x}}(t)-\widetilde{\mathbf{x}}_{e};\,\mathbf{h}(t))\right)^{-1}\frac{\partial}{\partial\mathbf{h}}F(\widetilde{\mathbf{x}}(t)-\widetilde{\mathbf{x}}_{e};\,\mathbf{h}(t))\\
\mathbf{I}
\end{array}\right]\triangleq\left[\begin{array}{c}
\widehat{\varphi}(\widetilde{\mathbf{x}}(t);\,\mathbf{h}(t))\\
\mathbf{I}
\end{array}\right].\label{eqn:Phi(he)-hat}
\end{equation}
Note that the joint state $\widetilde{\mathbf{z}}_{e}=(\widetilde{\mathbf{x}}_{e},\,\mathbf{h}_{e})$
and $\widetilde{\mathbf{x}}(t)$ are the algorithm outputs at the
$t$-th algorithm iteration in \eqref{eqn:alg_proposed-with-compensation-1}-\eqref{eqn:alg_proposed-with compensation-2}.
The estimate $\widehat{\Phi}(\widetilde{\mathbf{z}}_{e})$ in \eqref{eqn:Phi(he)-hat}
is quite accurate when the algorithm trajectory $\widetilde{\mathbf{x}}(t)$
is close enough to the equilibrium $\widetilde{\mathbf{x}}^{*}(\mathbf{h}(t))$
(i.e. $\|\widetilde{\mathbf{x}}_{e}\|$ is small enough). As a result,
the corresponding primal-dual algorithm iterations with compensation
is given by 
\begin{eqnarray}
\dot{\mathbf{x}} & = & \frac{d\mathbf{x}}{dt}=\left[\frac{\partial}{\partial\mathbf{x}}\mathcal{L}(\mathbf{x},\,\mathbf{\mathbf{\lambda}};\,\mathbf{h}(t))+\widehat{\varphi}_{\mathbf{x}}(\mathbf{x},\,\mathbf{\lambda};\,\mathbf{h}(t))^{T}\dot{\mathbf{h}}(t)\right]^{T}\label{eqn:alg_proposed-with-compensation-1}\\
\dot{\mathbf{\lambda}} & = & \frac{d\mathbf{\lambda}}{dt}=\left[\left(-\frac{\partial}{\partial\mathbf{\lambda}}\mathcal{L}(\mathbf{x},\,\mathbf{\mathbf{\lambda}};\,\mathbf{h}(t))+\widehat{\varphi}_{\mathbf{\lambda}}(\mathbf{x},\,\mathbf{\lambda};\,\mathbf{h}(t))^{T}\dot{\mathbf{h}}(t)\right)^{T}\right]_{\lambda}^{+}\label{eqn:alg_proposed-with compensation-2}
\end{eqnarray}
where $\widehat{\varphi}_{\mathbf{x}}(\mathbf{x},\,\mathbf{\lambda};\,\mathbf{h}(t))$
and $\widehat{\varphi}_{\mathbf{\lambda}}(\mathbf{x},\,\mathbf{\lambda};\,\mathbf{h}(t))$
are the primal and dual parts of the compensation term $\widehat{\varphi}(\widetilde{\mathbf{x}};\,\mathbf{h}(t))$. 

\begin{figure}
\begin{centering}
\includegraphics[scale=0.5]{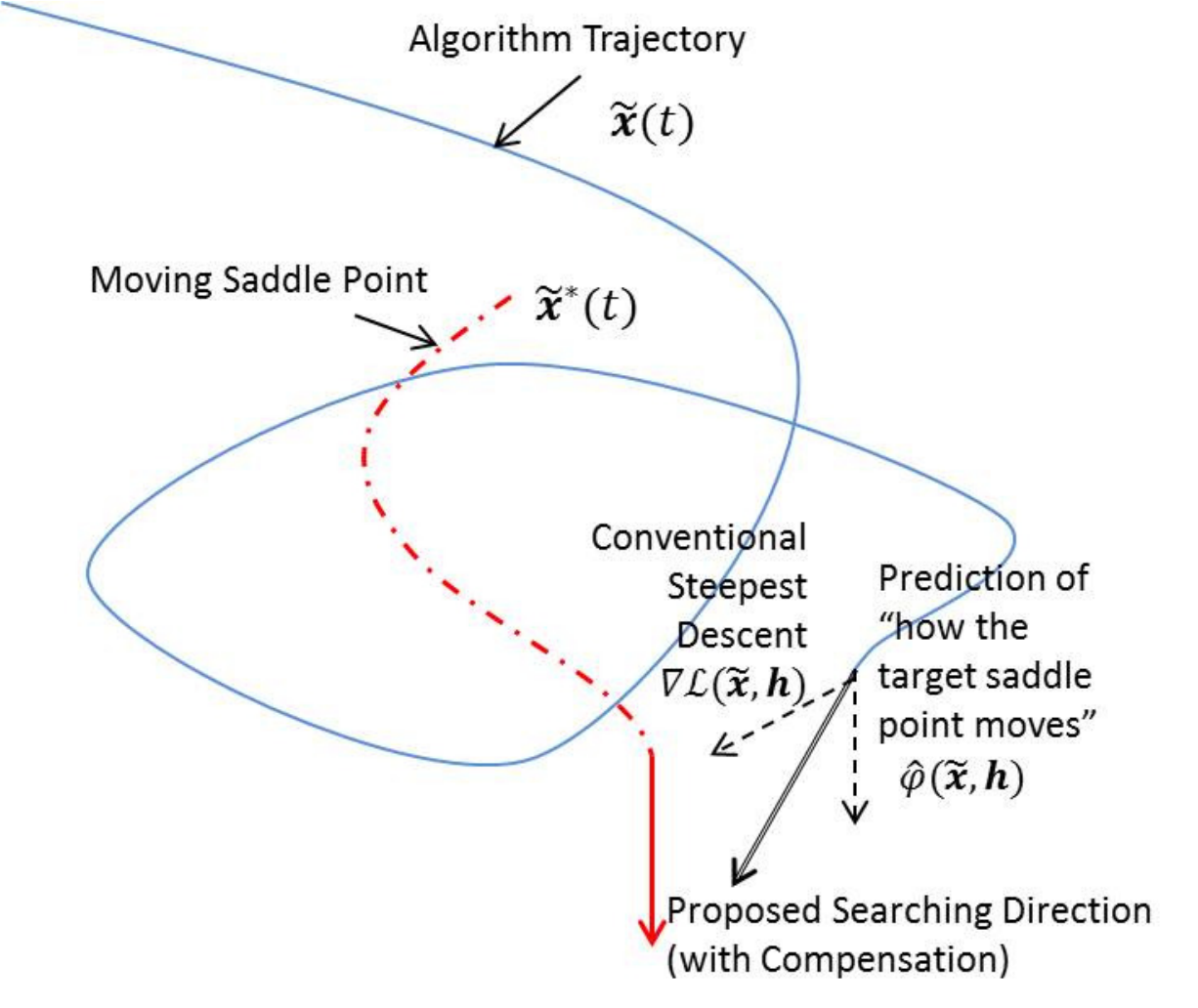}
\par\end{centering}

\begin{centering}
\caption{\label{fig:prediction-proposed alg}An illustration of the proposed
algorithm with adaptive compensation to improve the convergence behavior
of the primal-dual iterations in time-varying channels. Instead of
searching via the steepest descent direction $\nabla\mathcal{L}(\widetilde{\mathbf{x}},\mathbf{h})$
in the conventional primal-dual algorithm, it searches in a compensated
direction $\nabla\mathcal{L}(\mathbf{\widetilde{x}},\,\mathbf{h})+\widehat{\varphi}(\widetilde{\mathbf{x}},\,\mathbf{h})$
to offset the potential movement of the saddle point $\widetilde{\mathbf{x}}^{*}(t)$. }

\par\end{centering}

\end{figure}

The compensation terms in \eqref{eqn:alg_proposed-with-compensation-1}-\eqref{eqn:alg_proposed-with compensation-2}
can also be interpreted as a predictor on where the \emph{saddle point
}$\widetilde{\mathbf{x}}^{*}(\mathbf{h)}$ moves as illustrated in
Fig. \ref{fig:prediction-proposed alg}. We summarize the performance
of the proposed algorithm in the following theorem.
\begin{lyxThmQED}
\label{thm:Stability-of-z_e(u)-prosd-alg}(\emph{Tracking Performance
of the Compensated Primal-Dual Iteration for Strongly Concave-Convex
Saddle Point Problems):} Suppose that the compensation term $\widehat{\varphi}(\widetilde{\mathbf{x}};\,\mathbf{h}(t))$
is Lipschitz continuous on $\widetilde{\mathbf{x}}$ satisfying $\|\widehat{\varphi}(\widetilde{\mathbf{x}}(t);\,\mathbf{h}(t))-\widehat{\varphi}(\widetilde{\mathbf{x}}^{*}(t);\,\mathbf{h}(t))\|\leq L\|\widetilde{\mathbf{x}}_{e}(t)\|$
for all $\mathbf{h}(t)\in\mathcal{H}\subseteq\mathbb{R}^{q}$ and
$\lambda(t)\in\mathbb{R}_{+}^{m}$, and $\overline{\|u(t)\|}=\lim_{T\to\infty}\frac{1}{T}\int_{0}^{T}\|u(t)\|dt=\beta<\frac{a_{3}}{a_{4}L}$.
Then the average tracking error $\overline{\|\widetilde{\mathbf{x}}_{e}\|^{2}}$
of the proposed algorithm \eqref{eqn:alg_proposed-with-compensation-1}-\eqref{eqn:alg_proposed-with compensation-2}
converges to 0.\end{lyxThmQED}
\begin{proof}
Please refer to Appendix \ref{sec:App_Proof-of-Theorem_stability_prosd_alg}
for the proof.
\end{proof}

Similarly, for a degraded saddle point problem, the same proposed
algorithm \eqref{eqn:alg_proposed-with-compensation-1}-\eqref{eqn:alg_proposed-with compensation-2}
can be applied, and the degraded virtual dynamic system $\mathcal{Z}_{e}(\widehat{u})$
is given by 
\begin{equation}
\mathcal{Z}_{e}(\widehat{u}):\quad\dot{\mathbf{z}}_{e}=\mathcal{Z}_{e}(\mathbf{z}_{e})+\left(\Phi_{x}(\mathbf{h}_{e})-\widehat{\Phi}_{x}(\widetilde{\mathbf{z}}_{e})\right)u(t)\label{sys:alg-partial-z_e(u)}
\end{equation}
where 
\[
\widehat{\Phi}_{x}(\widetilde{\mathbf{z}}_{e})=\left[\begin{array}{c}
\widehat{\varphi}_{\mathbf{x}}(\mathbf{x},\,\mathbf{\lambda};\,\mathbf{h}(t))\\
\mathbf{I}
\end{array}\right]
\]
and the associated performance is summarized below.
\begin{lyxCorQED}
\label{cor:Stability-of-partial-z_e(u)-prosd-alg}(\emph{Tracking
Performance of the Compensated Primal-Dual Iteration for Degraded
Saddle Point Problems):} Suppose the compensation term $\widehat{\varphi}_{\mathbf{x}}(\mathbf{x},\,\mathbf{\lambda};\,\mathbf{h}(t))$
is Lipschitz continuous on $\mathbf{x}$ satisfying $\|\widehat{\varphi}_{\mathbf{x}}(\mathbf{x}(t),\,\mathbf{\lambda}(t);\,\mathbf{h}(t)-\widehat{\varphi}_{\mathbf{x}}(\mathbf{x}^{*}(t),\,\mathbf{\lambda}(t);\,\mathbf{h}(t))\|\leq L_{x}\|\mathbf{x}_{e}(t)\|$
for all $\mathbf{h}(t)\in\mathcal{H}\subseteq\mathbb{R}^{q}$ and
$\lambda(t)\in\mathbb{R}_{+}^{m}$, and $\overline{\|u(t)\|}=\lim_{T\to\infty}\frac{1}{T}\int_{0}^{T}\|u(t)\|dt=\beta<\frac{a_{3}}{a_{4}L_{x}}$.
Then the average tracking error $\overline{\|\mathbf{x}_{e}\|^{2}}$
of the proposed algorithm \eqref{eqn:alg_proposed-with-compensation-1}-\eqref{eqn:alg_proposed-with compensation-2}
converges to 0.
\end{lyxCorQED}

Corollary \ref{cor:Stability-of-partial-z_e(u)-prosd-alg} can be
obtained from Theorem \ref{thm:Stability-of-z_e(u)-prosd-alg} in
a straight-forward manner by replacing the variable $\widetilde{\mathbf{z}}_{e}$
with $\mathbf{z}_{e}$.

\subsection{Distributive Implementation}

The proposed algorithm \eqref{eqn:alg_proposed-with-compensation-1}-\eqref{eqn:alg_proposed-with compensation-2}
discussed in the previous section requires centralized computation
for the compensation term $\widehat{\varphi}(\widetilde{\mathbf{x}};\,\mathbf{h}(t))$.
However, for a wireless network, distributive solutions are usually
desired. One way for distributive implementation is via message passing
(i.e. communicating with other nodes to obtain non-local variables),
while each node computes a version of $\widehat{\varphi}(\widetilde{\mathbf{x}};\,\mathbf{h}(t))$
based on the global knowledge. However, this involves a significant
communication overhead and consumes a lot of computing resources.
In the following, we propose a low complexity distributive implementation
for computing the compensation term.

Consider an ad hoc network topology as depicted in Problem \ref{pro:Network-Utility-Maximization}
in Section \ref{subsub:App_NUM}. To facilitate distributive implementation
of the compensated primal-dual algorithm \eqref{eqn:alg_proposed-with-compensation-1}-\eqref{eqn:alg_proposed-with compensation-2},
we partition the whole network into $G$ groups of nodes and partition
the collection of optimization variables $\widetilde{\mathbf{x}}(t)=[\widetilde{\mathbf{x}}_{1}(t)|\widetilde{\mathbf{x}}_{2}(t)|\dots|\widetilde{\mathbf{x}}_{G}(t)]$
accordingly. Let $B(t)=\frac{\partial}{\partial\mathbf{\widetilde{x}}}F(\widetilde{\mathbf{x}}(t);\,\mathbf{h}(t))$
and $K(t)=\frac{\partial}{\partial\mathbf{h}}F(\widetilde{\mathbf{x}}(t);\,\mathbf{h}(t))$.
We then impose a block diagonal structure to the matrix $B(t)$ and
partition $K(t)$ as 
\begin{equation}
\widehat{B}(t)=\mbox{blkdiag}\{B_{\widetilde{\mathbf{x}}_{1}}(t),\, B_{\widetilde{\mathbf{x}}_{2}}(t),\dots,B_{\widetilde{\mathbf{x}}_{G}}(t)\},\quad\widehat{K}(t)^{T}=[K_{\widetilde{\mathbf{x}}_{1}}^{T}(t)\,|\, K_{\widetilde{\mathbf{x}}_{2}}^{T}(t)\,|\,\dots\,|\, K_{\widetilde{\mathbf{x}}_{G}}^{T}(t)]\label{eqn:prediction term - partition}
\end{equation}
where $\widetilde{\mathbf{x}}_{i}(t),\forall i=1,\dots,G$ is the
collection of variables for the $i$-th group of nodes, $B_{\widetilde{\mathbf{x}}_{i}}(t)=\frac{\partial}{\partial\mathbf{\widetilde{x}}_{i}}F_{\widetilde{\mathbf{x}}_{i}}(\widetilde{\mathbf{x}}(t);\,\mathbf{h}(t))$
is a block of $B(t)$ for the corresponding variables $\widetilde{\mathbf{x}}_{i}(t)$,
and $K_{\widetilde{\mathbf{x}}_{i}}(t)=\frac{\partial}{\partial\mathbf{h}}F_{\widetilde{\mathbf{x}}_{i}}(\widetilde{\mathbf{x}}(t);\,\mathbf{h}(t))$
is a block of $K(t)$ for the corresponding variables $\widetilde{\mathbf{x}}_{i}(t)$.
Therefore, the compensation term for the variable $\widetilde{\mathbf{x}}_{i}(t)$
is just $\widehat{\varphi}_{\widetilde{\mathbf{x}}_{i}}(t)=-B_{\widetilde{\mathbf{x}}_{i}}^{-1}(t)K_{\widetilde{\mathbf{x}}_{i}}(t)$.
For example, for a NUM problem with a topology depicted in Fig. \ref{fig:Multihop wireless network},
the network can be partitioned into two parts according to the two
links, where $\widetilde{\mathbf{x}}_{1}=\{r_{12},\, r_{13},\,\bm{\lambda}_{1}\}$
and $\widetilde{\mathbf{x}}_{2}=\{r_{23},\,\bm{\lambda}_{2}\}$ (i.e.
$\bm{\lambda}_{i}$ are the corresponding dual variables). The distributive
update equation for the collection of variable $\widetilde{\mathbf{x}}_{i}$
is 
\begin{equation}
\dot{\widetilde{\mathbf{x}}}_{i}=\left[\left(\nabla_{\widetilde{\mathbf{x}}_{i}}\mathcal{L}(\widetilde{\mathbf{x}};\,\mathbf{h}(t))+\widehat{\varphi}_{\widetilde{\mathbf{x}}_{i}}(\widetilde{\mathbf{x}};\,\mathbf{h}(t))\right)^{T}\right]_{\widetilde{\mathbf{x}}_{i}}^{+}\quad\forall i=1,\,\dots,\, G\label{eqn:alg-proposed-distributive}
\end{equation}
where $\nabla_{\widetilde{\mathbf{x}}_{i}}\mathcal{L}(\centerdot)$
denotes the searching direction by the gradient of $\frac{\partial\mathcal{L}(\centerdot)}{\partial\widetilde{\mathbf{x}}_{i}}$
and $[\centerdot]_{\widetilde{\mathbf{x}}_{i}}^{+}$ denotes the corresponding
projection for the variable $\widetilde{\mathbf{x}}_{i}$.

As the terms $B_{\widetilde{\mathbf{x}}_{i}}(t)$ and $K_{\widetilde{\mathbf{x}}_{i}}(t)$
just involve local variables, $\widehat{\varphi}_{\widetilde{\mathbf{x}}_{i}}(t)$
can be computed locally using only local information and hence the
update equation \eqref{eqn:alg-proposed-distributive} can be implemented
distributively. Note that in wireless communication networks, nodes
that are far away from each others usually have weak connections,
and hence $B(t)$ is actually a sparse matrix with dense diagonal
entries. Therefore, using the block diagonal version $\widehat{B}(t)$
in \eqref{eqn:prediction term - partition} will not contributes too
much performance loss in the proposed algorithm.

\section{\label{sec:Results-and-Discussions}Results and Discussions}

In this section, we shall simulate the tracking performance of the
primal-dual algorithms for \textcolor{black}{various saddle point
problems we studied before}. We evaluate the performance of the proposed
algorithms with the adaptive compensation term compared with the baseline
algorithms in \cite{Feijer:2010uq,Nedic2009,Kallio:1994fk}. \textcolor{black}{The
system performance loss due to the time varying channels and the tracking
errors is also illustrated. }

\subsection{Convergence Performance of the Strongly Concave-convex Saddle Point
Problem and the Degraded Saddle Point Problem}

\begin{figure}
\begin{centering}
\includegraphics[scale=0.55]{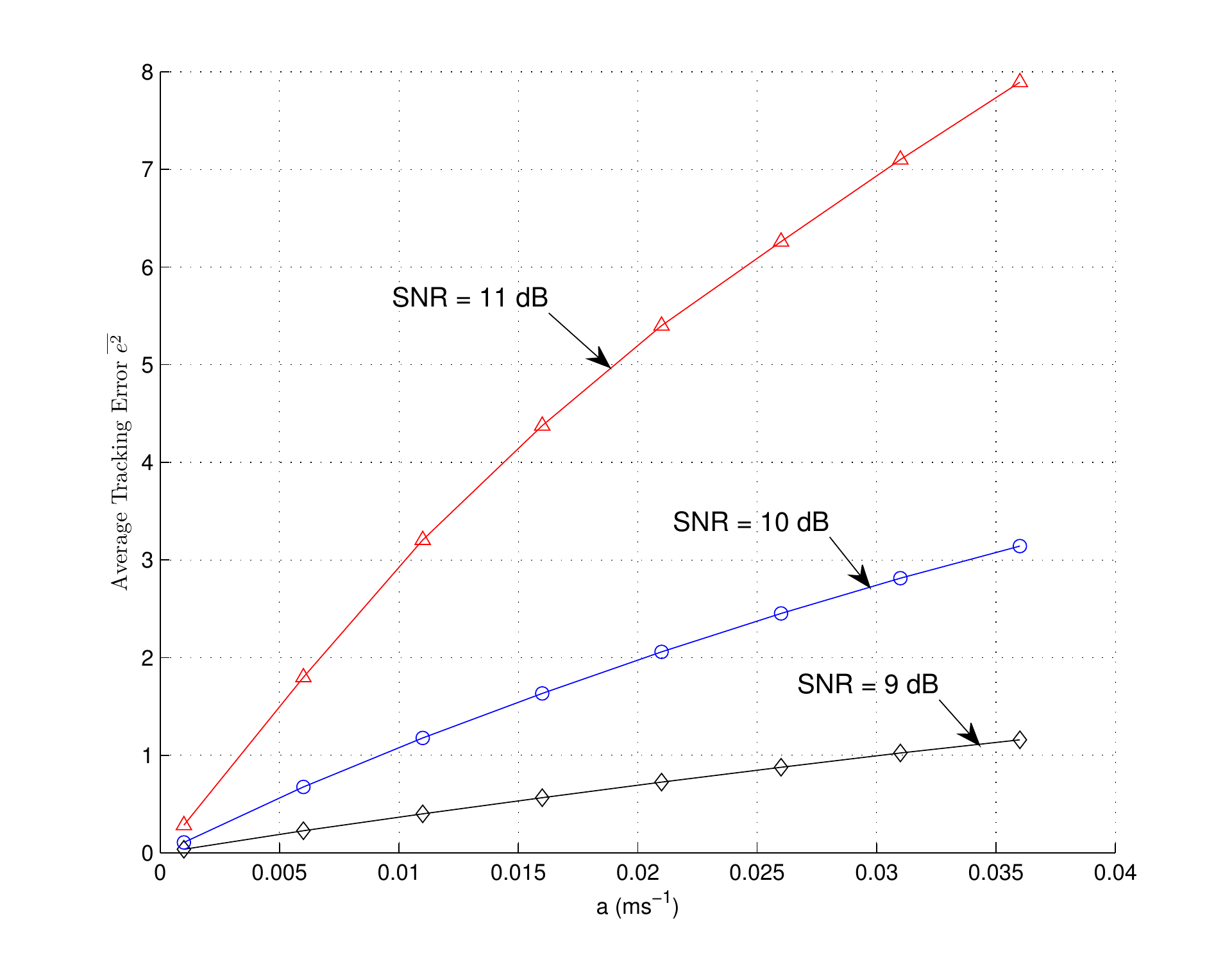}
\par\end{centering}

\caption{\label{fig:TrackErr_matrix_Q}The convergence performance of the primal-dual
algorithm for the strongly concave-convex saddle point problem given
in Section \ref{sub:Numerical-Example_JammingGame}. The CSI model
is given by $\dot{\mathbf{h}}=A(\mathbf{h}-\bar{\mathbf{h}})+\sqrt{2a}w(t)$,
where $A=-aI$, $\bar{\mathbf{h}}=\mathbf{1}$, and $w(t)$ is a zero-mean
unit-variance white Gaussian process. The variance of the CSI $\mathbf{h}(t)$
is normalized to unity, and the parameter $a$ controls the time-correlation
of the CSI. It shows that as the SNR increases, the tracking error
also increases for the same CSI variation rate in terms of $a$.}
\end{figure}

\begin{figure}
\begin{centering}
\includegraphics[scale=0.55]{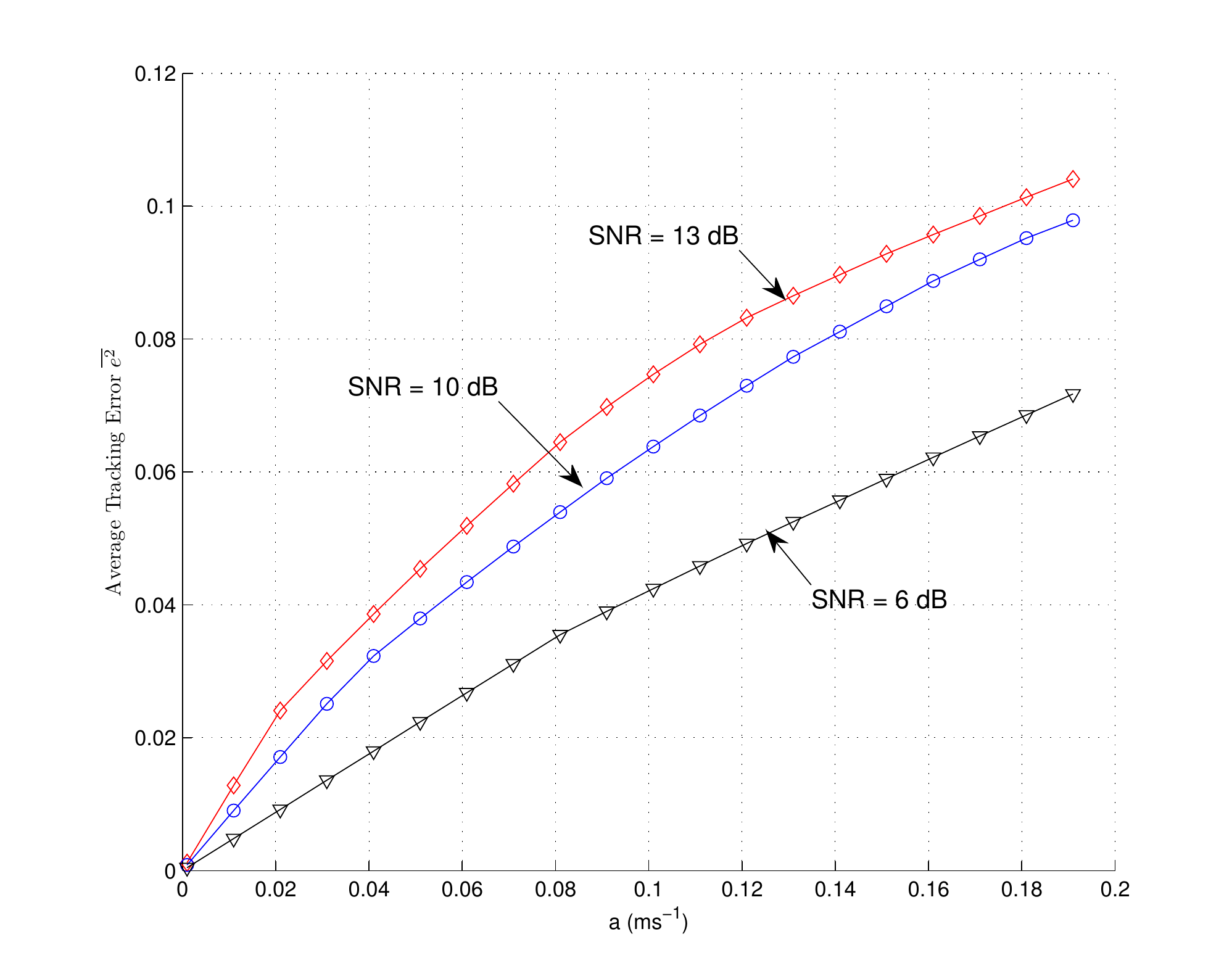}
\par\end{centering}

\caption{\label{fig:TrackErr_NUM}The convergence performance of the primal-dual
algorithm for the degraded saddle point problem given in Section \ref{sub:Numerical example - NUM}.
The CSI model is given by $\dot{\mathbf{h}}=-a(\mathbf{h}-\bar{\mathbf{h}})+\sqrt{2a}w(t)$,
where $\bar{\mathbf{h}}=\mathbf{1}$ and $w(t)$ is a zero-mean unit-variance
white Gaussian process. The variance of the CSI $\mathbf{h}(t)$ is
normalized to unity, and the parameter $a$ controls the time-correlation
of the CSI. It shows that as the SNR increases, the tracking error
also increases for the same CSI variation rate in terms of $a$.}

\end{figure}

The simulations of the convergence performance for the strongly concave-convex
saddle point problem \textcolor{black}{and the degraded saddle point
problem are} based on the examples described in Section \ref{sub:Numerical-Example_JammingGame}
\textcolor{black}{and Section \ref{sub:Numerical example - NUM},
respectively}. \textcolor{black}{The average tracking error in the
jamming problem is defined as $\overline{e^{2}}=\mathbb{E}\left[\|\widetilde{\mathbf{x}}(t)-\widetilde{\mathbf{x}}^{*}(t)\|^{2}\right]$
where $\widetilde{\mathbf{x}}(t)=\mbox{vec}([Q(t)\; Z(t)])$, while
$\overline{e^{2}}=\mathbb{E}\left[\|\mathbf{r}(t)-\mathbf{r}^{*}(t)\|^{2}\right]$
defines the average tracking error in the NUM problem, where $\mathbf{r}(t)$
is the rate allocation variables. In both cases, the} CSI model is
specified by an AR process $\dot{\mathbf{h}}=-a(\mathbf{h}(t)-\bar{\mathbf{h}})+\sqrt{2a}w(t)$,
where $\bar{\mathbf{h}}=\bm{1}$, $w(t)$ is a zero-mean unity-variance
Gaussian process and $a$ is a parameter to synthesize the fading
rate. \textcolor{black}{Fig. \ref{fig:TrackErr_matrix_Q} and Fig.\ref{fig:TrackErr_NUM}
illustrate the average tracking errors $\overline{e^{2}}$ versus
the fading rate parameter $a$ in both of the problems.} As $a$
increases, the CSI changes more rapidly and the tracking errors of
the primal dual algorithm increase. Moreover, for the same parameter
$a$, the larger the average receiving SNR is, the larger tracking
error will be.

\subsection{Convergence Performance of the Proposed Algorithm with Compensation}

\begin{figure}
\begin{centering}
\includegraphics[scale=0.16]{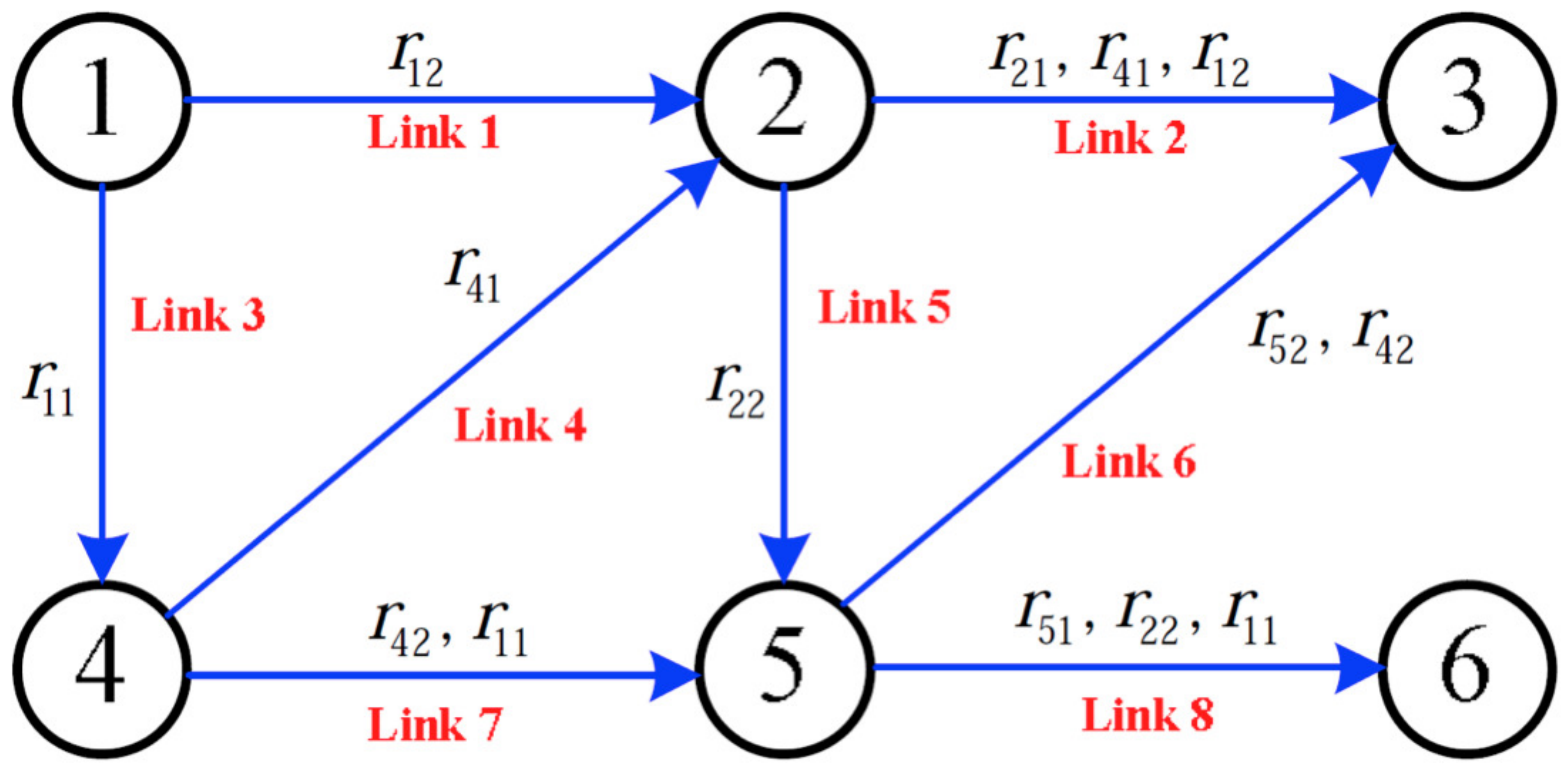}
\par\end{centering}

\caption{\label{fig:Multihop-network-6nodes}\textcolor{black}{A specific
example of wireless ad hoc network with 6 nodes 8 directed links and
8 data flows. The data flows  are delivered simutaneously with fixed
routes. Links from the same transmitting node occupy different subbands
and do not interfere with each other. The interference at each receiving
node is handled by multiuser detection (MUD) techniques. Allowable
transmission rate for each link is determined by the capacity region
at the corresponding receiving node.}}

\end{figure}

\textcolor{black}{To evaluate the algorithm performance in a more
concrete communication scenario, we consider a NUM problem of a wireless
ad hoc network with 6 nodes 8 links and 8 data flows as illustrated
in Fig.\ref{fig:Multihop-network-6nodes}. The data flows are delivered
simultaneously with fixed routes. Links from the same transmitting
node occupy different subbands while the interference at each receiving
node is handled by multiuser detection (MUD) techniques. The problem
is to optimize the transmit power under the time varying channels
to support required traffic data rates. Similar ad hoc network scenario
is also considered in \cite{ElBatt2004}. Similarly, the CSI model
is specified by the AR process $\dot{\mathbf{h}}=-a(\mathbf{h}(t)-\bar{\mathbf{h}})+\sqrt{2a}w(t)$,
where $a$ is a parameter to determine the CSI fading rate.}

\textcolor{black}{The proposed algorithms with adaptive compensations
are compared with the following baseline algorithms. }
\begin{itemize}
\item \textbf{Baseline 1 - Conventional Primal-Dual Gradient Algorithm (ConPDGA)}
\cite{Arrow1958,Feijer:2010uq}: The primal and dual variables are
updated simultaneously in \eqref{eqn:alg_Arrow-gradient-h(t)-1}-\eqref{eqn:alg_Arrow-gradient-h(t)-2}
in the gradients of the objective function.
\item \textbf{Baseline 2 - Averaging Primal-Dual Gradient Algorithm (AvgPDGA)}
\cite{Nedic2009}: The primal and dual variables are updated in the
gradients of the objective function, and the approximate primal solutions
are generated by averaging over the past primal solutions.
\item \textbf{Baseline 3 - Perturbed Primal-Dual Gradient Algorithm (PerPDGA)}
\cite{Kallio:1994fk,Kallio:1994uq}:\textbf{ }The primal and dual
variables are updated in the gradients evaluated at perturbed points
that are generated from current points via auxiliary mappings.
\item \textbf{Proposed Primal-Dual Algorithm - Centralized:} The primal
and dual variables are updated in a compensated direction in \eqref{eqn:alg_proposed-with-compensation-1}-\eqref{eqn:alg_proposed-with compensation-2}
to offset the potential movement of the saddle point.
\item \textbf{Proposed Primal-Dual Algorithm - Distributed:} The primal
and dual variables are updated in a compensated direction in \eqref{eqn:alg-proposed-distributive}
with imposed block diagonal structure in \eqref{eqn:prediction term - partition}.
\end{itemize}

\begin{figure}
\begin{centering}
\includegraphics[scale=0.53]{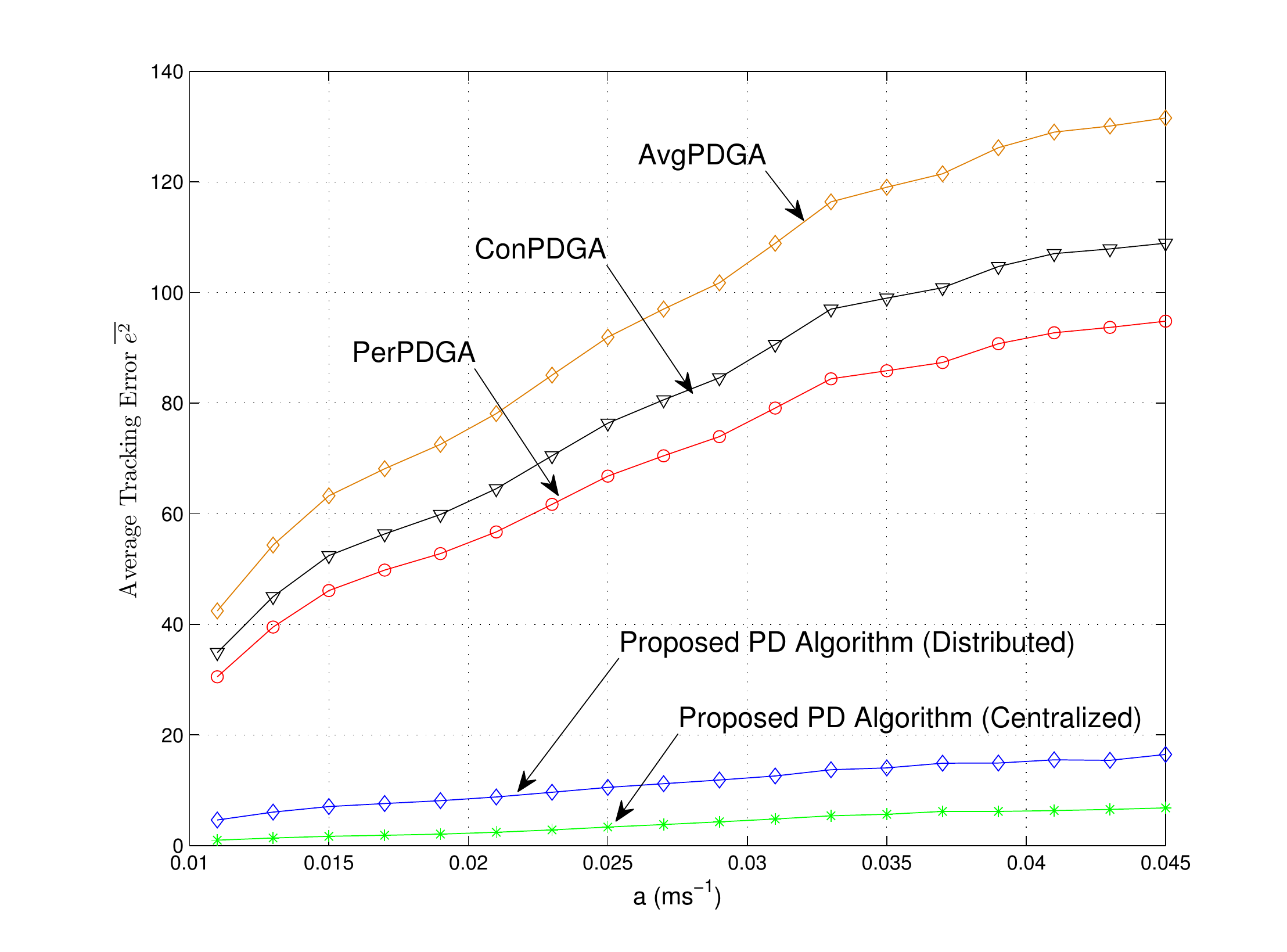}
\par\end{centering}

\caption{\label{fig:TrackingErr_NUM_comp}\textcolor{black}{The convergence
performance comparison of the proposed algorithms and the baseline
algorithms\cite{Feijer:2010uq,Kallio:1994fk,Nedic2009}, which solve
a NUM under the wireless network in Fig.\ref{fig:Multihop-network-6nodes}.
The CSI model is given by $\dot{\mathbf{h}}=-a(\mathbf{h}-\bar{\mathbf{h}})+\sqrt{2a}w(t)$,
where $\bar{\mathbf{h}}=\mathbf{1}$ and $w(t)$ is a zero-mean unit-variance
white Gaussian process. The variance of the CSI $\mathbf{h}(t)$ is
normalized to unity, and the parameter $a$ controls the time-correlation
of the CSI. The tracking errors of the proposed algorithms are much
smaller than those of the baselines.}}

\end{figure}

Fig. \ref{fig:TrackingErr_NUM_comp} shows the comparison of the average
tracking error $\overline{e^{2}}$ versus $a$ in the time varying
CSI model. The tracking errors have been reduced greatly after introducing
the compensation terms in both of the proposed algorithms.

\subsection{\textcolor{black}{Performance Loss due to Tracking Errors}}

\textcolor{black}{In this section, we consider the degradation of
the system performance (network throughput) and the CSI fading rate.
Due to the time varying channels, the problem constraints (e.g. the
link capacity constraints) may not be satisfied at every time slot.
As a result, packet drops may occur and we define the system performance
as the \emph{average actual throughput} given by: 
\[
\bar{R}=\mathbb{E}\left[\sum_{k\in\mathcal{K}}\left(\sum_{(s,d)\in\mathcal{T}(k)}r_{sd}\right)\times1\left\{ \tilde{\mathbf{c}}_{k}\in\mathcal{C}_{k}(\mathbf{p};\mathbf{h})\right\} \right]
\]
where $\mathcal{K}$ denotes the set of receiving nodes, $\mathcal{T}(k)$
denotes the set of traffic flows arrived at node $k$, $\tilde{\mathbf{c}}_{k}$
denotes the vector of link transmission rates of the links towards
node $k$, $\mathcal{C}_{k}(\centerdot)$ denotes the capacity region
at receiving node $k$ and $1(\centerdot)$ is the indicator function
to account for the penalty of constraint violation due to time varying
CSI. For example, at node $k=5$ in Fig. \ref{fig:Multihop-network-6nodes},
the set of traffic flows arrived is $\mathcal{T}(5)=\{r_{22,}r_{42,}r_{11}\}$
and $\tilde{\mathbf{c}}_{5}=(\tilde{r}_{5},\tilde{r}_{7})$ is a rate
vector corresponding to transmission rates $\tilde{r}_{5}$ and $\tilde{r}_{7}$
allocated to link 5 and link 7, respectively.}

\begin{figure}
\begin{centering}
\includegraphics[scale=0.53]{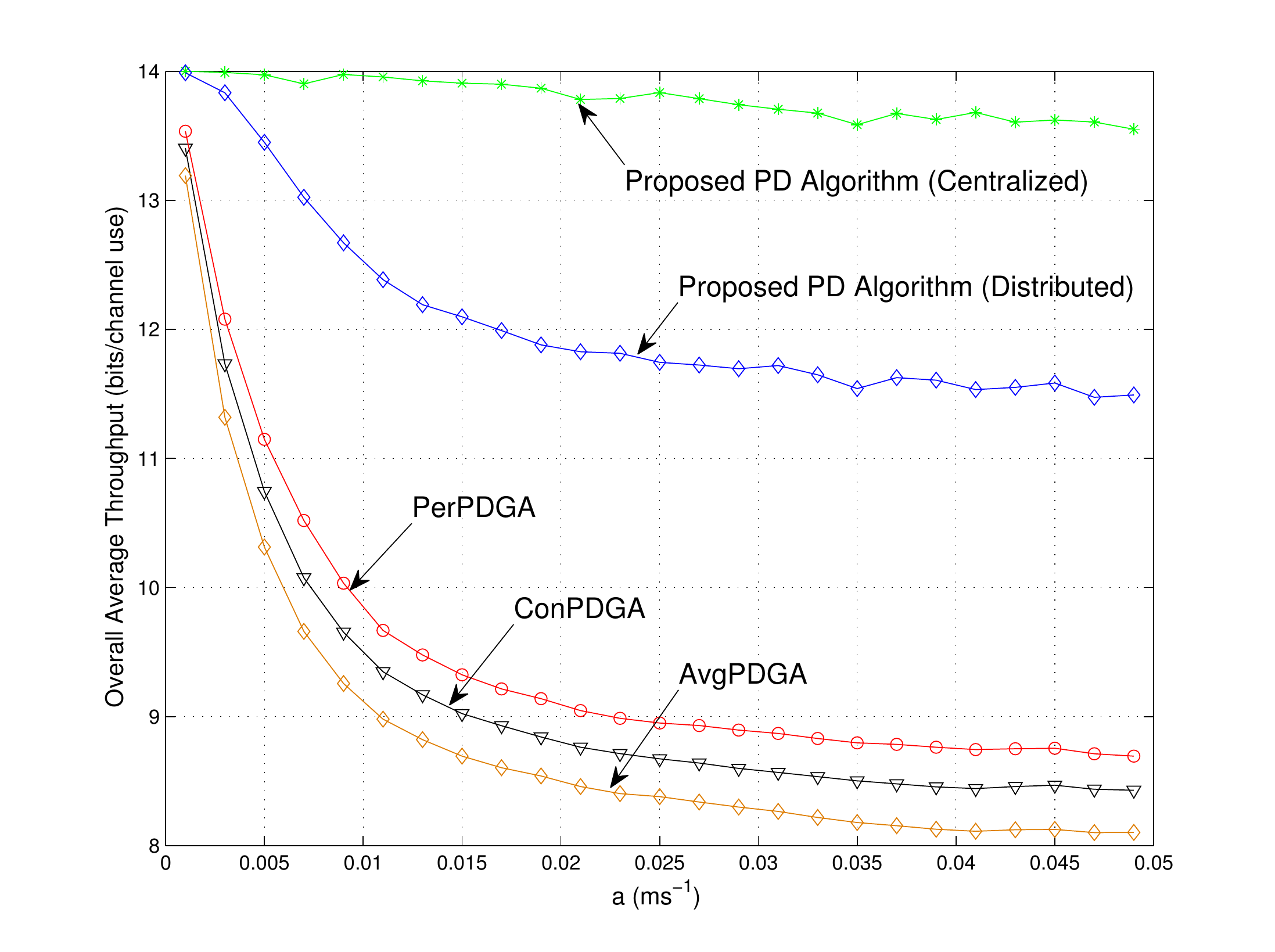}
\par\end{centering}

\caption{\label{fig:Throughput}\textcolor{black}{Average network throughput
versus the channel fading rate parameter $a$, while applying the
various proposed and baseline algorithms to solve the NUM problem
under the network in Fig.\ref{fig:Multihop-network-6nodes}. The average
throughput decreases when the fading rate $a$ increases. The proposed
algorithms have much higher network throughput than all the other
baselines. }}

\end{figure}

\textcolor{black}{Fig.\ref{fig:Throughput} shows the the average
network throughput versus the channel fading rate $a$. The average
throughput decreases when the fading rate $a$ increases, as indicated
by a growing average tracking error in Fig.\ref{fig:TrackingErr_NUM_comp}.
The results also show that the proposed algorithms significantly outperform
over all the other baselines.}

\section{Conclusions\label{sec:Conclusions}}

In this paper, we have analyzed the convergence behavior of the primal-dual
algorithm for solving a saddle point problem under time varying CSI,
which was modeled as an AR process. The convergence results have been
derived by studying the stabilities of the equivalent virtual dynamic
systems based on the Lyapunov theory from the control theoretical
approach. We showed that for both the strongly concave-convex saddle
point problem and the degraded saddle point problem, the average tracking
errors were given by $\mathcal{O}(\alpha^{2})$, where $\alpha^{2}$
represents the average power of the exogenous excitation induced to
the CSI dynamics. Based on these analyses, we have proposed a novel
adaptive primal-dual algorithm with a predictive compensation to counteract
the effects of the time varying CSI. We showed that the average tracking
error of the proposed algorithm converges to zero despite time varying
CSI. Numerical results were consistent with our analysis and the proposed
algorithm demonstrated significantly better convergence performance
over the baseline schemes.

\appendices

\section{Proof of Theorem \ref{thm:Connections-between-Virtual-Saddle point}\label{sec:App_Proof-of-Theorem_connection}}
\begin{proof}
The equilibrium point $\widetilde{\mathbf{x}}^{*}(\mathbf{h}(t))$
of the virtual dynamic system $\widetilde{\mathcal{X}}$ in \eqref{eqn:def-virtual dynamic system}
satisfies $\widetilde{f}(\widetilde{\mathbf{x}}^{*}(\mathbf{h}(t));\,\mathbf{h}(t))=0$.
By Definition \ref{def:Virtual-Dynamic-System}, it is equivalent
to 
\begin{equation}
\frac{\partial}{\partial\mathbf{x}}\mathcal{L}(\mathbf{x}^{*},\,\mathbf{\mathbf{\lambda}}^{*};\,\mathbf{h}(t))=0,\quad\lambda_{i}\frac{\partial}{\partial\mathbf{\lambda}_{i}}\mathcal{L}(\mathbf{x}^{*},\,\mathbf{\mathbf{\lambda}}^{*};\,\mathbf{h}(t))=0,\;\forall i\label{eqn:app-saddle point optimality cond}
\end{equation}
which is the optimality condition for the saddle point problem. As
the saddle point problem \eqref{eqn:min-max} is strongly concave
in $\mathbf{x}$, the saddle point $\left[\mathbf{x}^{*}(\mathbf{h}(t)),\,\mathbf{\lambda^{*}(}\mathbf{h}(t))\right]$
is uniquely determined by \eqref{eqn:app-saddle point optimality cond},
which proofs that the saddle point and equilibrium are identical. 

From Definition \ref{def:Virtual-Dynamic-System}, the virtual dynamic
system \eqref{eqn:def-virtual dynamic system} and the primal-dual
algorithm \eqref{eqn:alg_Arrow-gradient-h(t)-1}-\eqref{eqn:alg_Arrow-gradient-h(t)-2}
share the same formulation. Hence the trajectories of the two are
the same.
\end{proof}

\section{Proof of Lemma \ref{lem:stability of x_e with time-invariant CSI}\label{sec:App_Proof-of-Lemma_stability_x_e}}
\begin{proof}
Take $V(\mathbf{\widetilde{x}}_{e})=\frac{1}{2\kappa}\mathbf{\widetilde{x}}_{e}^{T}\mathbf{\widetilde{x}}_{e}$
as a Lyapunov function defined along the trajectory of the virtual
dynamic system \eqref{sys:error-dynamic-system}. Suppose that $\widetilde{\mathbf{x}}_{e}+\widetilde{\mathbf{x}}^{*}$
is in the interior of the domain $\mathbb{R}^{n}\times\mathbb{R}_{+}^{m}$.
From  
\[
\nabla\widetilde{f}_{e}(\mathbf{\widetilde{x}}_{e};\,\mathbf{h})=\kappa\left[\begin{array}{cc}
\nabla_{\mathbf{x}}^{2}\mathcal{L} & \nabla_{\mathbf{x}\lambda}\mathcal{L}\\
-\nabla_{\mathbf{x}\lambda}^{T}\mathcal{L} & -\nabla_{\lambda}^{2}\mathcal{L}
\end{array}\right]
\]
and the strong convexity / concavity properties $\nabla_{\mathbf{x}}^{2}\mathcal{L}\preceq-M_{x}I$
and $\nabla_{\mathbf{\lambda}}^{2}\mathcal{L}\succeq M_{\lambda}I$,
we obtain 
\[
\mathbf{\widetilde{x}}_{e}^{T}\nabla\widetilde{f}_{e}(\mathbf{\widetilde{x}}_{e};\,\mathbf{h})\mathbf{\widetilde{x}}_{e}\leq-\kappa\min\{M_{x},\, M_{\lambda}\}\|\mathbf{\widetilde{x}}_{e}\|^{2}\triangleq-\kappa M\|\mathbf{\widetilde{x}}_{e}\|^{2}
\]
Then $\dot{V}(\mathbf{0})=0$ and for any $\mathbf{\widetilde{x}}_{e}\neq0$,
\begin{eqnarray*}
\dot{V}(\mathbf{\widetilde{x}}_{e}) & = & \nabla V(\widetilde{\mathbf{x}}_{e})\dot{\mathbf{\widetilde{x}}}_{e}=\frac{1}{\kappa}\mathbf{\widetilde{x}}_{e}^{T}\dot{\mathbf{\widetilde{x}}_{e}}=\frac{1}{\kappa}\mathbf{\widetilde{x}}_{e}^{T}\widetilde{f}_{e}(\mathbf{\widetilde{x}}_{e};\,\mathbf{h})\\
 & = & \frac{1}{\kappa}\mathbf{\widetilde{x}}_{e}^{T}\widetilde{f}_{e}(\mathbf{0}_{+};\,\mathbf{h})+\frac{1}{\kappa}\mathbf{\widetilde{x}}_{e}^{T}\int_{0}^{1}\nabla\widetilde{f}_{e}(\xi\mathbf{\widetilde{x}}_{e};\,\mathbf{h})d\xi\mathbf{\widetilde{x}}_{e}\\
 & \leq & \frac{1}{\kappa}\mathbf{\widetilde{x}}_{e}^{T}\widetilde{f}_{e}(\mathbf{0}_{+};\,\mathbf{h})-\int_{0}^{1}\frac{1}{\xi}Md\|\xi\mathbf{\widetilde{x}}_{e}\|^{2}\\
 & \leq & -2M\|\mathbf{\widetilde{x}}_{e}\|^{2}
\end{eqnarray*}
where $\mathbf{\widetilde{x}}_{e}^{T}\widetilde{f}_{e}(\mathbf{0}_{+};\,\mathbf{h})=<\widetilde{\mathbf{x}}-\widetilde{\mathbf{x}}^{*},\nabla\bar{\mathcal{L}}(\widetilde{\mathbf{x}}^{*};\mathbf{h})>\leq0$
from the optimality condition of the original saddle point problem.
Here for simplicity, we denote $\nabla\bar{\mathcal{L}}(\widetilde{\mathbf{x}};\mathbf{h})=[\frac{\partial}{\partial\mathbf{x}}\mathcal{L}(\mathbf{x},\mathbf{\lambda};\mathbf{h})\;-\frac{\partial}{\partial\lambda}\mathcal{L}(\mathbf{x},\mathbf{\lambda};\mathbf{h})]^{T}$.

Therefore, using the Lyapunov theory \cite{Khalil1996}, we conclude
that the virtual dynamic system is exponentially stable for every
time-invariant $\mathbf{h}$.
\end{proof}

\section{Proof of Lemma \ref{lem:Stability-z_e-under Quasi-Time Varying}\label{sec:App_Proof-of-Lemma_stability z_e quasi time varying}}
\begin{proof}
One possible choice of Lyapunov function on virtual state $\widetilde{\mathbf{z}}_{e}$
along the algorithm trajectory is $V(\widetilde{\mathbf{z}}_{e})=\frac{1}{2\kappa}\widetilde{\mathbf{x}}_{e}^{T}\widetilde{\mathbf{x}}_{e}+\frac{1}{2}\mathbf{h}_{e}^{T}\mathbf{h}_{e}$.
Then we have 
\[
\min\{\frac{1}{2\kappa},\frac{1}{2}\}\|\widetilde{\mathbf{z}}_{e}\|^{2}\leq V(\widetilde{\mathbf{z}}_{e})=\frac{1}{2\kappa}\|\widetilde{\mathbf{x}}_{e}\|^{2}+\frac{1}{2}\|\mathbf{h}_{e}\|^{2}\leq\max\{\frac{1}{2\kappa},\,\frac{1}{2}\}\|\widetilde{\mathbf{z}}_{e}\|^{2}
\]
and 
\begin{eqnarray*}
\dot{V}(\widetilde{\mathbf{z}}_{e}) & = & \frac{\partial V}{\partial\widetilde{\mathbf{z}}_{e}}Z(\widetilde{\mathbf{z}}_{e})=\left[\begin{array}{cc}
\frac{1}{\kappa}\widetilde{\mathbf{x}}_{e}^{T} & \mathbf{h}_{e}^{T}\end{array}\right]\left[\begin{array}{c}
\widetilde{f}_{e}(\widetilde{\mathbf{x}}_{e};\,\mathbf{h}_{e}+\bar{\mathbf{h}})+\varphi(\mathbf{h}_{e}+\bar{\mathbf{h}})A\mathbf{h}_{e}\\
A\mathbf{h}_{e}
\end{array}\right]\\
 & = & \frac{1}{\kappa}\widetilde{\mathbf{x}}_{e}\widetilde{f}_{e}(\widetilde{\mathbf{x}}_{e};\,\mathbf{h})+\mathbf{h}^{T}A\mathbf{h}+\frac{1}{\kappa}\widetilde{\mathbf{x}}_{e}\varphi(\mathbf{h})A\mathbf{h}\\
 & \leq & -2M\|\widetilde{\mathbf{x}}_{e}\|^{2}+\lambda_{\max}(A)\|\mathbf{h}\|^{2}+\frac{1}{\kappa}\|\widetilde{\mathbf{x}}_{e}\|\|\varphi(\mathbf{h})A\|\|\mathbf{h}\|\\
 & \leq & -\left[\min\{2M,\,-\lambda_{\max}(A)\}-\frac{1}{\kappa}\|\varphi(\mathbf{h})A\|\right]\|\widetilde{\mathbf{z}}_{e}\|^{2}\\
 & = & -a_{3}\|\widetilde{\mathbf{z}}_{e}\|^{2}
\end{eqnarray*}
where $a_{3}\triangleq\min\{2M,\,-\lambda_{\max}(A)\}-\frac{1}{\kappa}\|\varphi(\mathbf{h})A\|>0$
by the condition in \eqref{eqn:lem-stability-time-varying_cond1_a3}.
In addition, we have 
\[
\left\Vert \frac{\partial V}{\partial\widetilde{\mathbf{z}}_{e}}\right\Vert =\left\Vert \left[\begin{array}{cc}
\frac{1}{\kappa}\widetilde{\mathbf{x}}_{e}^{T} & \mathbf{h}^{T}\end{array}\right]\right\Vert \leq\sqrt{\frac{1}{\kappa^{2}}\|\widetilde{\mathbf{x}}_{e}\|^{2}+\|\mathbf{h}\|^{2}}\leq\max\{\frac{1}{\kappa},1\}\|\widetilde{\mathbf{z}}_{e}\|
\]
Therefore, using Lyapunov theory \cite{Khalil1996}, we conclude that
the joint system is exponentially stable and \eqref{eqn:lem-stability-time-varying_V}-\eqref{eqn:lem-stability-time-varying_pV}
are satisfied with $a_{1}=\min\{1/2\kappa,\,1/2\}$, $a_{2}=\max\{1/2\kappa,\,1/2\}$,
$a_{3}=\min\{2M,\,-\lambda_{\max}(A)\}-\frac{1}{\kappa}\|\varphi(\mathbf{h})A\|$
and $a_{4}=\max\{1/\kappa,1\}$. 
\end{proof}

\section{Proof of Theorem \ref{thm:Stablity of Z_e(u)}\label{sec:App_Proof-of-Theorem_stability_z_e}}
\begin{proof}
From Lemma \ref{lem:Stability-z_e-under Quasi-Time Varying}, there
exists a Lyapunov function $V(\widetilde{\mathbf{z}}_{e})$ defined
on the joint state $\widetilde{\mathbf{z}}_{e}$ along the trajectory
$\mathbf{\dot{\widetilde{z}}}_{e}=\widetilde{\mathcal{Z}}(\widetilde{\mathbf{z}}_{e})$,
satisfying \eqref{eqn:lem-stability-time-varying_V}-\eqref{eqn:lem-stability-time-varying_pV}.
Then, for the system \eqref{sys:z-u(t)}, 
\begin{eqnarray*}
\dot{V}(\widetilde{\mathbf{z}}_{e}) & = & \frac{\partial V}{\partial\widetilde{\mathbf{z}}_{e}}\widetilde{\mathcal{Z}}(\widetilde{\mathbf{z}}_{e})+\frac{\partial V}{\partial\widetilde{\mathbf{z}}_{e}}\Phi(\mathbf{h}_{e})u(t)\leq-a_{3}\|\widetilde{\mathbf{z}}_{e}\|^{2}+a_{4}\|\Phi(\mathbf{h}_{e})\|\|\widetilde{\mathbf{z}}_{e}\|\|u(t)\|
\end{eqnarray*}

Suppose there exists $\epsilon>0$, such that 
\begin{equation}
\lim_{T\to\infty}\frac{1}{T}\int_{0}^{T}\|\widetilde{\mathbf{z}}_{e}(\tau)\|^{2}d\tau\geq\frac{a_{4}^{2}}{a_{3}^{2}}\gamma^{2}\alpha^{2}+\epsilon\label{eqn:proof-thm2-assumption-1}
\end{equation}
Then given any $t>0$, there exists $T_{0}(t)>0$, such that for all
$T>T_{0}$, 
\[
\frac{1}{t+T}\int_{0}^{t+T}\|\widetilde{\mathbf{z}}_{e}(\tau)\|^{2}d\tau\geq\frac{a_{4}^{2}}{a_{3}^{2}}\gamma^{2}\alpha^{2}+\frac{\epsilon}{2}
\]
 and there exists $T_{1}(t)>0$, such that for all $T>T_{1},$ 
\[
\frac{1}{t+T}\int_{0}^{t}\|\widetilde{\mathbf{z}}_{e}(\tau)\|^{2}d\tau\leq\frac{\epsilon}{4}.
\]
Choose $0<\eta<\frac{\epsilon}{4}\frac{a_{3}^{2}}{a_{4}^{2}\gamma^{2}}$,
and then from $\overline{\|u(t)\|^{2}}=\lim_{T\to\infty}\frac{1}{T}\int_{0}^{T}\|u(t)\|^{2}dt\leq\alpha^{2}$,
there exists $T_{2}(t)>0$, such that for all $T>T_{2}$, 
\begin{equation}
\frac{1}{T}\int_{t}^{t+T}\|u(\tau)\|^{2}d\tau=\frac{t+T}{T}\frac{1}{t+T}\int_{0}^{t+T}\|u(\tau)\|^{2}d\tau\leq\frac{t+T}{T}\left(\alpha^{2}+\eta\right)\label{eqn:proof-thm2-arg-3}
\end{equation}
Finally, let $T_{3}(t)=\max\{T_{0}(t),T_{1}(t),T_{2}(t)\}$. Then
for all $T>T_{3}$, 
\begin{equation}
\frac{1}{T}\int_{t}^{t+T}\|\widetilde{\mathbf{z}}_{e}(\tau)\|^{2}d\tau=\frac{t+T}{T}\left[\frac{1}{t+T}\left(\int_{0}^{t+T}\|\widetilde{\mathbf{z}}_{e}(\tau)\|^{2}d\tau-\int_{0}^{t}\|\widetilde{\mathbf{z}}_{e}(\tau)\|^{2}d\tau\right)\right]\geq\frac{t+T}{T}\left(\frac{a_{4}^{2}}{a_{3}^{2}}\gamma^{2}\alpha^{2}+\frac{\epsilon}{4}\right)\label{eqn:proof-thm2-arg-4}
\end{equation}
Hence the $T$-step Lyapunov drift is given by 
\begin{eqnarray*}
 &  & V\left(\widetilde{\mathbf{z}}_{e}(t+T)\right)-V(\widetilde{\mathbf{z}}_{e}(t))\\
 & \leq & -\int_{t}^{t+T}a_{3}\|\widetilde{\mathbf{z}}_{e}(\tau)\|^{2}d\tau+\int_{t}^{t+T}a_{4}\gamma\|\widetilde{\mathbf{z}}_{e}(\tau)\|\|u(\tau)\|d\tau\\
 & \leq & -a_{3}\int_{t}^{t+T}\|\widetilde{\mathbf{z}}_{e}(\tau)\|^{2}d\tau+\sqrt{\int_{t}^{t+T}a_{4}^{2}\gamma^{2}\|\widetilde{\mathbf{z}}_{e}(\tau)\|^{2}d\tau}\centerdot\sqrt{\int_{t}^{t+T}\|u(\tau)\|^{2}d\tau}\\
 & = & \left(-a_{3}\sqrt{\int_{t}^{t+T}\|\widetilde{\mathbf{z}}_{e}(\tau)\|^{2}d\tau}+a_{4}\gamma\sqrt{\int_{t}^{t+T}\|u(\tau)\|^{2}d\tau}\right)\sqrt{\int_{t}^{t+T}\|\widetilde{\mathbf{z}}_{e}(\tau)\|^{2}dt\tau}\\
 & \leq & \left(-a_{3}\sqrt{\left(\frac{a_{4}^{2}}{a_{3}^{2}}\gamma^{2}\alpha^{2}+\frac{\epsilon}{4}\right)(t+T)}+a_{4}\gamma\sqrt{\left(\alpha^{2}+\eta\right)(t+T)}\right)\sqrt{\int_{t}^{t+T}\|\widetilde{\mathbf{z}}_{e}(\tau)\|^{2}dt\tau}\\
 & < & 0
\end{eqnarray*}
by \eqref{eqn:proof-thm2-arg-3} and \eqref{eqn:proof-thm2-arg-4}.
By the property of the Lyapunov function $V(\widetilde{\mathbf{z}}_{e})$
in \eqref{eqn:lem-stability-time-varying_V} , for any given $t>0$,
we have $\|\widetilde{\mathbf{z}}_{e}(t+T)\|<\|\widetilde{\mathbf{z}}_{e}(t)\|$
for all $T>T_{3}$. Hence, we have $\lim_{t\to\infty}\|\widetilde{\mathbf{z}}_{e}(t)\|=0$,
which contradicts to the assumption in \eqref{eqn:proof-thm2-assumption-1}.
So the average trajectory $\overline{\|\widetilde{\mathbf{z}}_{e}\|^{2}}\leq\frac{a_{4}^{2}}{a_{3}^{2}}\gamma^{2}\alpha^{2}$,
which proofs the stability.
\end{proof}

\section{Proof of Lemma \ref{lem:Partial-Exponential-Stability of z_e quasi-time}
\label{sec:App_Proof-of-Lemma_partial stability z_e quasi}}
\begin{proof}
A simple choice of Lyapunov function is $V(\mathbf{z}_{e})=V_{0}(\mathbf{x}_{e})+\frac{1}{2}\mathbf{h}_{e}^{T}\mathbf{h}_{e}$,
where $V_{0}(\mathbf{x}_{e})=V_{x}(\widetilde{\mathbf{x}}_{e})\big|_{\lambda_{e}=0}=\frac{1}{2\kappa}\mathbf{x}_{e}^{T}\mathbf{x}_{e}$
and $\frac{\partial V_{0}(\mathbf{x}_{e})}{\partial\widetilde{\mathbf{x}}_{e}}=\frac{1}{\kappa}[\mathbf{x}_{e}^{T}\;\mathbf{0}^{T}]$.
Hence we have 
\[
\min\{\frac{1}{2\kappa},\,\frac{1}{2}\}\|\mathbf{z}_{e}\|^{2}\leq c_{1}\|\mathbf{x}_{e}\|^{2}+\frac{1}{2}\|\mathbf{h}\|^{2}\leq V(\mathbf{z}_{e})\leq c_{2}\|\mathbf{x}_{e}\|^{2}+\frac{1}{2}\|\mathbf{h}\|^{2}\leq\max\{\frac{1}{2\kappa},\,\frac{1}{2}\}\|\mathbf{z}_{e}\|^{2}
\]
and 
\begin{eqnarray*}
\frac{dV(\mathbf{z}_{e})}{dt} & = & \frac{\partial V(\mathbf{z}_{e})}{\partial\widetilde{\mathbf{z}}_{e}}\frac{d\widetilde{\mathbf{z}}_{e}}{dt}=\left[\begin{array}{cc}
\frac{\partial V_{0}(\mathbf{x}_{e})}{\partial\widetilde{\mathbf{x}}_{e}} & \mathbf{h}_{e}^{T}\end{array}\right]\left[\begin{array}{c}
\widetilde{f}_{e}(\widetilde{\mathbf{x}}_{e};\,\mathbf{h}_{e}+\bar{\mathbf{h}})+\varphi(\mathbf{h}_{e}+\bar{\mathbf{h}})A\mathbf{h}_{e}\\
A\mathbf{h}_{e}
\end{array}\right]\\
 & = & \frac{\partial V_{0}(\mathbf{x}_{e})}{\partial\widetilde{\mathbf{x}}_{e}}\widetilde{f}_{e}(\widetilde{\mathbf{x}}_{e};\,\mathbf{h}_{e}+\bar{\mathbf{h}})+\frac{\partial V_{0}(\mathbf{x}_{e})}{\partial\widetilde{\mathbf{x}}_{e}}A\mathbf{h}_{e}+\mathbf{h}_{e}^{T}A\mathbf{h}_{e}\\
 & \leq & -c_{3}\|\mathbf{x}_{e}\|^{2}+c_{4}\|\mathbf{x}_{e}\|\|\varphi_{x}(\mathbf{h})A\|\|\mathbf{h}_{e}\|+\lambda_{\max}(A)\|\mathbf{h}_{e}\|^{2}\\
 & \leq & -\left[\min\{c_{3},\,-\lambda_{\max}(A)\}-c_{4}\|\varphi_{x}(\mathbf{h})A\|\right]\|\mathbf{z}_{e}\|^{2}\\
 & \triangleq & -a_{3}\|\mathbf{z}_{e}\|^{2}
\end{eqnarray*}
where $c_{3}=2M_{x}$ and $c_{4}=1/\kappa$ following the discussion
in Lemma \ref{lem:Partial Exponential Stability of x_e}. Moreover,
\[
\left\Vert \frac{\partial V(\mathbf{z}_{e})}{\partial\mathbf{z}_{e}}\right\Vert =\left\Vert \left[\begin{array}{cc}
\frac{\partial V_{0}(\mathbf{x}_{e})}{\partial\widetilde{\mathbf{x}}_{e}} & \mathbf{h}^{T}\end{array}\right]\right\Vert \leq\sqrt{c_{4}^{2}\|\mathbf{x}_{e}\|^{2}+\|\mathbf{h}\|^{2}}\leq\sqrt{\max\{\frac{1}{\kappa^{2}},\,1\}}\|\mathbf{z}_{e}\|
\]
Therefore, using the Lyapunov theory \cite{Khalil1996}, we conclude
that the system $\dot{\mathbf{z}}_{e}=\mathcal{Z}(\mathbf{z}_{e})$
is exponentially stable for the joint state $\mathbf{z}_{e}=(\mathbf{x}_{e},\,\mathbf{h})$
and \eqref{eqn:lem-partial-exp_z_V}-\eqref{eqn:lem-partial-exp_z_pV}
are satisfied with $a_{1}=\min\{1/(2\kappa),\,1/2\}$, $a_{2}=\max\{1/(2\kappa),\,1/2\}$,
$a_{3}=\min\{2M_{x},\,-\lambda_{\max}(A)\}-\frac{1}{\kappa}\|\varphi_{x}(\mathbf{h})A\|$
and $a_{4}=\sqrt{\max\{1/\kappa^{2},\,1\}}$. 
\end{proof}

\section{Proof of Theorem \ref{thm:Stability-of-z_e(u)-prosd-alg}\label{sec:App_Proof-of-Theorem_stability_prosd_alg}}
\begin{proof}
Consider the virtual dynamic system with compensation in \eqref{sys:alg_z_u(t)}.
Since 
\[
\Phi(\mathbf{h}_{e})-\widehat{\Phi}(\mathbf{\widetilde{z}}_{e})=\left[\begin{array}{c}
\widehat{\varphi}(\widetilde{\mathbf{x}}^{*};\,\mathbf{h}(t))-\widehat{\varphi}(\widetilde{\mathbf{x}};\,\mathbf{h}(t))\\
0
\end{array}\right]
\]
then $\|\Phi(\mathbf{h}_{e})-\widehat{\Phi}(\mathbf{\widetilde{z}}_{e})\|=\|\widehat{\varphi}(\widetilde{\mathbf{x}}^{*};\,\mathbf{h}(t))-\widehat{\varphi}(\widetilde{\mathbf{x}};\,\mathbf{h}(t))\|\leq L\|\widetilde{\mathbf{x}}_{e}\|\leq L\|\widetilde{\mathbf{z}}_{e}\|$.
Applying the Lyapunov function\textbf{ }$V(\widetilde{\mathbf{z}}_{e})$
in Lemma \ref{lem:Stability-z_e-under Quasi-Time Varying} to the
virtual dynamic system $\widetilde{\mathcal{Z}}_{e}(\widehat{u})$,
we obtain 
\begin{eqnarray*}
\dot{V}(\widetilde{\mathbf{z}}_{e}) & = & \frac{\partial V}{\partial\widetilde{\mathbf{z}}_{e}}\widetilde{\mathcal{Z}}(\widetilde{\mathbf{z}}_{e})+\frac{\partial V}{\partial\widetilde{\mathbf{z}}_{e}}\Phi(\mathbf{h}_{e})u(t)\leq-a_{3}\|\widetilde{\mathbf{z}}_{e}\|^{2}+a_{4}\|\Phi(\mathbf{h}_{e})-\widehat{\Phi}(\mathbf{\widetilde{z}}_{e})\|\|u(t)\|\|\widetilde{\mathbf{z}}_{e}\|\\
 & \leq & -a_{3}\|\widetilde{\mathbf{z}}_{e}\|^{2}+a_{4}L\|u(t)\|\|\widetilde{\mathbf{z}}_{e}\|^{2}
\end{eqnarray*}

Suppose there exists $\epsilon>0$, such that 
\begin{equation}
\lim_{T\to\infty}\frac{1}{T}\int_{t}^{t+T}\|\widetilde{\mathbf{z}}_{e}(\tau)\|^{2}d\tau>\epsilon\label{eqn:proof-thm4-assumption-1}
\end{equation}
Then there exists $T_{0}(t)>0$, such that for all $T>T_{0}$, $\frac{1}{T}\int_{t}^{t+T}\|\widetilde{\mathbf{z}}_{e}(\tau)\|^{2}d\tau>\frac{\epsilon}{2}$. 

As $\lim_{T\to\infty}\frac{1}{T}\int_{t}^{t+T}\|u(\tau)\|d\tau=\beta<\frac{a_{3}}{a_{4}L}$,
we choose $\eta=\left(\frac{a_{3}}{a_{4}L}-\beta\right)/2$. Then
there exists $T_{1}(t)>0$, such that for all $T>T_{1}$, $\frac{1}{T}\int_{t}^{t+T}\|u(\tau)\|d\tau<\beta+\eta<\frac{a_{3}}{a_{4}L}$. 

Observed that, $\widetilde{\mathbf{z}}_{e}(t)$ only depends on all
the realizations before $u(t_{-})$. Therefore $\|u(t)\|$ and $\|\widetilde{\mathbf{z}}_{e}(t)\|^{2}$
are independent and hence, 
\[
\lim_{T\to\infty}\frac{1}{T}\int_{t}^{t+T}\|u(\tau)\|\|\widetilde{\mathbf{z}}_{e}(\tau)\|^{2}d\tau=\lim_{T\to\infty}\frac{1}{T}\int_{t}^{t+T}\|u(\tau)\|d\tau\centerdot\frac{1}{T}\int_{t}^{t+T}\|\widetilde{\mathbf{z}}_{e}(\tau)\|^{2}d\tau.
\]
Therefore, if we choose any $0<\delta<\eta\epsilon/2$, then there
exists $T_{2}(t)>0$, such that for all $T>T_{2}$, 
\[
\frac{1}{T}\int_{t}^{t+T}\|u(\tau)\|\|\widetilde{\mathbf{z}}_{e}(\tau)\|^{2}d\tau\leq\frac{1}{T}\int_{t}^{t+T}\|u(\tau)\|d\tau\centerdot\frac{1}{T}\int_{t}^{t+T}\|\widetilde{\mathbf{z}}_{e}(\tau)\|^{2}d\tau+\delta.
\]
As a result, let $T_{3}(t)=\max\{T_{0}(t),T_{1}(t),T_{2}(t)\}$, and
then for all $T>T_{3},$ the $T$-step Lyapunov drift becomes 
\begin{eqnarray*}
V(\widetilde{\mathbf{z}}_{e}(t+T))-V(\widetilde{\mathbf{z}}_{e}(t)) & \leq & -\int_{t}^{t+T}a_{3}\|\widetilde{\mathbf{z}}_{e}(\tau)\|^{2}d\tau+a_{4}L\int_{t}^{t+T}\|u(\tau)\|\|\widetilde{\mathbf{z}}_{e}(\tau)\|^{2}d\tau\\
 & \leq & -\int_{t}^{t+T}a_{3}\|\widetilde{\mathbf{z}}_{e}(\tau)\|^{2}d\tau+a_{4}L\left(\frac{1}{T}\int_{t}^{t+T}\|u(\tau)\|d\tau\int_{t}^{t+T}\|\widetilde{\mathbf{z}}_{e}(\tau)\|^{2}d\tau+\delta T\right)\\
 & \leq & \left(-a_{3}+a_{4}L\centerdot\frac{1}{T}\int_{t}^{t+T}\|u(\tau)\|d\tau\right)\int_{t}^{t+T}\|\widetilde{\mathbf{z}}_{e}(\tau)\|^{2}d\tau+a_{4}L\delta T\\
 & \leq & -a_{4}L\eta\int_{t}^{t+T}\|\widetilde{\mathbf{z}}_{e}(\tau)\|^{2}d\tau+a_{4}L\delta T\\
 & \leq & \left(-\eta\frac{\epsilon}{2}+\delta\right)a_{4}LT\\
 & < & 0
\end{eqnarray*}

By the property of the Lyapunov function $V(\widetilde{\mathbf{z}}_{e})$
in \eqref{eqn:lem-stability-time-varying_V}, given any $t>0$, we
have $\|\widetilde{\mathbf{z}}_{e}(t+T)\|<\|\widetilde{\mathbf{z}}_{e}(t)\|$
for all $T>T_{3}$. Hence, $\lim_{t\to\infty}\|\widetilde{\mathbf{z}}_{e}(t)\|=0$,
which contradicts to the assumption in \eqref{eqn:proof-thm4-assumption-1}.
So the average trajectory $\overline{\|\widetilde{\mathbf{z}}_{e}\|^{2}}$
converges to 0, and hence the tracking error $\overline{\|\widetilde{\mathbf{x}}_{e}\|^{2}}$
converges to 0.
\end{proof}
\bibliographystyle{IEEEtran}
\bibliography{Bib_Converge_Analysis}

\newpage

\end{document}